\documentclass[numberedappendix]{emulateapj}

\begin{document}  

\title{Candidate Disk Wide Binaries in the Sloan Digital Sky Survey}

\author{
Branimir Sesar\altaffilmark{1},
\v{Z}eljko Ivezi\'{c}\altaffilmark{1},
Mario Juri\'{c}\altaffilmark{2}
}

\altaffiltext{1}{University of Washington, Dept.~of Astronomy, Box
                           351580, Seattle, WA 98195-1580}
\altaffiltext{2}{Institute for Advanced Study, 1 Einstein Drive,
                           Princeton, NJ 08540}

\begin{abstract}

Using SDSS Data Release 6, we construct two independent samples of candidate
stellar wide binaries selected as i) pairs of unresolved sources with angular
separation in the range $3\arcsec - 16\arcsec$, ii) common proper motion pairs
with $5\arcsec - 30\arcsec$ angular separation, and make them publicly
available. These samples are dominated by disk stars, and we use them to
constrain the shape of the main-sequence photometric parallax relation
$M_r(r-i)$, and to study the properties of wide binary systems. We estimate
$M_r(r-i)$ by searching for a relation that minimizes the difference between
distance moduli of primary and secondary components of wide binary candidates.
We model $M_r(r-i)$ by a fourth degree polynomial and determine the coefficients
using Markov Chain Monte Carlo fitting, independently for each sample. Both
samples yield similar relations, with the largest systematic difference of 0.25
mag for F0 to M5 stars, and a root-mean-square scatter of 0.13 mag. A similar
level of agreement is obtained with photometric parallax relations recently
proposed by \citet{jur08}. The measurements show a root-mean-square scatter of
$\sim0.30$ mag around the best fit $M_r(r-i)$ relation, and a mildly
non-Gaussian distribution. We attribute this scatter to metallicity effects and
additional unresolved multiplicity of wide binary components. Aided by the
derived photometric parallax relation, we construct a series of high-quality
catalogs of candidate main-sequence binary stars. These range from a sample of
$\sim17,000$ candidates with the probability of each pair to be a physical
binary (the ``efficiency'') of $\sim65\%$, to a volume-limited sample of
$\sim1,800$ candidates with an efficiency of $\sim90\%$. Using these catalogs,
we study the distribution of semi-major axes of wide binaries, $a$, in the
$2,000 < a < 47,000$~AU range. We find the observations to be well described by
the \"Opik distribution, $f(a)\propto 1/a$, for $a<a_{break}$, where $a_{break}$
increases roughly linearly with the height $Z$ above the Galactic plane
($a_{break}\propto12,300\,Z{\rm [kpc]}^{0.7}$~AU). The number of wide binary
systems with $100 \, {\rm AU} < a < a_{break}$, as a fraction of the total
number of stars, decreases from 0.9\% at $Z=0.5$ kpc to 0.5\% at $Z=3$ kpc. The
probability for a star to be in a wide binary system is independent of its
color. Given this color, the companions of red components seem to be drawn
randomly from the stellar luminosity function, while blue components have a
larger blue-to-red companion ratio than expected from luminosity function.
\end{abstract}

\keywords{binaries: visual --- stars: distances --- Hertzsprung-Russell diagram}

\section{Introduction}
\label{introduction}

Binary systems can be roughly divided into close (semi-major axes $a\la10$ AU)
and wide (semi-major axes $a\ga100$ AU, \citealt{cha07}) pairs. Close binary
systems have long been recognized as useful tools for studies of stellar
properties. For example, the stellar parameters such as the masses and radii of
individual stars are readily determined to high confidence using eclipsing
binaries \citep{and91}. Wide binary systems have proven to be a tool for studies
of star formation processes, as well as an exceptionally useful tracer of local
potential and tidal fields through which they traverse. Specifically, they were
used to place the constraints on the nature of halo dark matter \citep{ycg04}
and to explore the dynamical history of the Galaxy \citep{aph07}. A further
comprehensive list of current applications of wide binaries can be found in
\citet{cha07}.

Close binaries, owing to their relatively short orbital periods and equally
short timescales of brightness or spectrum fluctuations, are fairly easy to
detect. Unambiguous identification of wide binary systems, on the other hand,
requires accurate astrometry on much longer timescales, as these systems have
orbital periods $\ga10,000$ years. However, instead of requiring unambiguous
identification, large samples of {\em candidate} wide binaries can be selected
by simply assuming that pairs of stars with small angular separation are also
gravitationally bound \citep{bs81,gou95}, or by searching for common proper
motion pairs \citep{luy79,pov94,aph00,gs03,cg04,lb07}. The angular separation
method is simple to apply, but it also introduces a relatively large number of
false candidates due to chance association of nearby pairs. The contamination by
random associations can be reduced by imposing constraints, such as the common
proper motion, or by requiring that the stars are at similar distances. The
distances can be inferred through a variety of means, one of which is the use of
an appropriate photometric parallax\footnote{Also known as ``color-luminosity
relation''.} relation.

The photometric parallax relation provides the absolute magnitude of a star
given that star's color and metallicity. There are a number of proposed
photometric parallax relations for main sequence stars in the literature that
differ in the methodology used to derive them, photometric systems, and the
absolute magnitude and metallicity range in which they are applicable. Not all
of them are mutually consistent, and most exhibit significant intrinsic scatter
of order a half a magnitude or more (see Figure~2 in
\citealt[hereafter J08]{jur08}).

Instead of using an existing relation to select wide binaries, we propose a
novel method that {\em simultaneously} derives the photometric parallax relation
{\em and} selects a sample of wide binary candidates. The method relies on the
fact that components of a physical binary have equal distance moduli
($m_1 - M_1 = m_2 - M_1$) and therefore
$\delta \equiv \Delta M - \Delta m \equiv (M_2 - M_1) - (m_2 - m_1)  = 0$.
Assuming that both stars are on the main sequence, and the {\em shape} of the
adopted photometric parallax relation is correct, the difference in absolute
magnitudes $\Delta M=M_2-M_1$ calculated from the parallax relation must equal
the measured difference of apparent magnitudes, $\Delta m=m_2-m_1$. The
$\Delta M = \Delta m$ equality for binaries must be valid irrespective of color,
and therefore represents a test of the validity of the adopted photometric
parallax relation or, alternatively, a way to estimate the parallax relation.

In practice, the distribution of $\delta$ will not be a delta-function both due
to instrumental (finite photometric precision) and physical effects (true
vs.~apparent pairs). However, for {\em true} wide binaries, the distribution of
$\delta$ is expected to be narrow, strongly peaked at zero, and the individual
$\delta$ values are expected to be uncorrelated with color. In contrast, the
distribution of $\delta$ values for randomly associated stellar pairs (hereafter
random pairs) should be much broader even when the correct photometric parallax
relation is adopted, reflecting the different distances of components of
projected binary pairs. This dichotomy can be used to assign a probability to
each candidate, of whether it is a true physical binary or a result of chance
projection on the sky.

The paper is organized as follows. In Section~\ref{data}, we give an overview
of the SDSS imaging data, and describe the selection, completeness and
population composition of two initial, independent samples of candidate
binaries. In Section~\ref{method} we describe the photometric parallax
estimation method, compare the best-fit photometric parallax relations to the
J08 relation, and analyze the scatter in predicted absolute magnitudes. The
properties of wide binaries, such as the color and spatial distributions, are
analyzed in Section~\ref{properties}. Finally, the results and their
implications for future surveys are discussed in Section~\ref{discussion}.

\section{The Data}
\label{data}

\subsection{Overview of the SDSS Imaging Data}
\label{sdss_overview}

Thanks to the quality of its photometry and astrometry, as well as the large
sky coverage, the SDSS stands out among available optical sky surveys. The SDSS
provides homogeneous and deep ($r<22.5$) photometry in five bandpasses ($u$,
$g$, $r$, $i$, and $z$, \citealt{gun98,hog02,smi02,gun06,tuc06}) accurate to
0.02 mag (rms scatter) for unresolved sources not limited by photon statistics
\citep{scr02,ive03}, and with a zeropoint uncertainty of 0.02 mag \citep{ive04}.
The survey sky coverage of 10,000 deg$^2$ in the northern Galactic cap and 300
deg$^2$ in the southern Galactic cap results in photometric measurements for
well over 100 million stars and a similar number of galaxies \citep{sto02}. The
recent Data Release 6 \citep{amc08} lists\footnote{See
\url[HREF]{http://www.sdss.org/dr6}} photometric data for 287 million unique
objects observed in 9583 deg$^2$ of sky, and can be accessed through the Catalog
Archive Server\footnote{\url[HREF]{http://cas.sdss.org}} (CAS) 
CasJobs\footnote{\url[HREF]{http://casjobs.sdss.org/CasJobs/}}
interface. Astrometric positions are accurate to better than $0.1\arcsec$ per
coordinate (rms) for sources with $r<20.5$ \citep{pie03}, and the morphological
information from the images allows reliable star-galaxy separation to
$r\sim21.5$ \citep{lup02}.

The five-band SDSS photometry can be used for very detailed source
classification, e.g., separation of quasars and stars \citep{ric02}, spectral
classification of stars to within one to two spectral subtypes
\citep{len98,fin00,haw02,cov07}, identification of horizontal-branch and RR
Lyrae stars \citep{yan00,sir04,ive05,ses07}, and low-metallicity G and K giants
\citep{hel03}.

Proper motion data exist for SDSS sources matched to the USNO-B1.0 catalog
\citep{mon03}. We take proper motion measurements from the \citet{mun04} catalog
based on astrometric measurements from the SDSS and Palomar Observatory Sky
Surveys (POSS-I; \citealt{ma63}, POSS-II; \citealt{rei91}). Despite the sizable
random and systematic astrometric errors in the Schmidt surveys, the combination
of a long baseline (50 years for the POSS-I survey) and a recalibration of the
photographic data using positions of SDSS galaxies, results in median random
errors for proper motions of only 3 mas yr$^{-1}$ for $r < 19.5$ (per
coordinate), with substantially smaller systematic errors \citep{mun04}.
Following a recommendation by Munn et al., when using their catalog we select
SDSS stars with only one USNO-B match within $1\arcsec$, and require proper
motion rms fit residuals to be less than 350 mas in both coordinates. We note
that the proper motion measurements publicly available as a part of SDSS Data
Release 6 are known to have significant systematic errors (Munn et al., 
in prep.). Here we use a revised set of proper motion measurements which will
become publicly available as a part of SDSS Data Release 7.

\subsection{The Initial Sample of Close Resolved Stellar Pairs}
\label{initial_select}

For objects in the SDSS catalog, the photometric pipeline \citep{lup02} sets a
number of flags that indicate the status of each object, warn of possible
problems with the image itself, and warn of possible problems in the measurement
of various quantities associated with the object. These flags can be used to
remove duplicate detections (in software) of the same object, and to select
samples of unresolved sources with good photometry.

According to the SDSS Catalog Archive Server ``Algorithms'' webpage\footnote{
\url[HREF]{http://cas.sdss.org/dr6/en/help/docs/algorithm.asp?key=flags}},
duplicate detections of the same objects can be removed by considering only
those which have the ``status'' flag set to PRIMARY. We consider only PRIMARY
objects, and select those with good photometry by requiring that the BINNED1
flag is set to 1, and PSF\_FLUX\_INTERP, DEBLEND\_NOPEAK, INTERP\_CENTER,
BAD\_COUNTS\_ERROR, NOTCHECKED, NOPROFILE, PEAKCENTER, and EDGE image processing
flags are set to 0 in the $gri$ bands. The moving unresolved sources, such as
asteroids, are avoided by selecting sources with the DEBLENDED\_AS\_MOVING flag
set to 0.

Good photometric accuracy (mean PSF magnitude errors $<0.03$ mag, see Figure~1
in \citealt{ses07}) is obtained by selecting sources with $14<r'<20.5$, where
$r'$ is the $r$ band PSF magnitude uncorrected for ISM extinction. The PSF
magnitudes corrected for ISM extinction (using maps by \citealt{SFD98}), and
used throughout this work, are noted as $u$, $g$, $r$, $i$, and $z$.

To create the initial sample of resolved stellar pairs, we
query\footnote{SQL queries are listed in Appendix~\ref{appendix}} the CAS
``Neighbors'' table (lists all SDSS pairs within $30\arcsec$) for pairs of
sources that pass the above criteria, and that have
\begin{equation}
(r_1-r_2)[(g-i)_1-(g-i)_2]>0\label{Teff},
\end{equation}
where the subscript 1 is hereafter assigned to the brighter component. With this
condition we require that the component with bluer $g-i$ color is brighter in
the $r$ band. About $40\%$ of random pairs are rejected with this condition. We
estimate that about $3\%$ of true binary systems might be excluded by this cut
(due to uncertainties in the $g-i$ color caused by photometric errors), but
their exclusion does not significantly influence our results.

We select $\sim4.2$ million pairs for the initial sample of resolved stellar
pairs, and plot the observed distribution of angular separation $\theta$,
$f_{obs}(\theta)$, in Figure~\ref{ang_sep} ({\em top}). For a uniform (random)
distribution of stars, the number of neighboring stars within an annulus
$\Delta \theta$ increases linearly with $\theta$, and therefore, the number
of random pairs also increases with $\theta$. To find the number of random pairs
as a function of $\theta$, we fit $f_{rnd}(\theta)= C \, \theta$ to the
$f_{obs}$ histogram (in the $\theta > 15\arcsec$ region), and find $C=9043$
arcsec$^{-1}$. For large separation angles ($\theta > 15\arcsec$) the two
distributions closely match, indicating that the majority of observed pairs are
simply random associations, and are not physically related. At separation
angles smaller than $\sim15\arcsec$ the frequency of observed pairs shows an
excess, suggesting the presence of true, gravitationally bound systems. However,
even at small separation angles, the selected pairs include a non-negligible
fraction of random pairs and require further refinement, or careful statistical
accounting for random contamination.

Throughout this work we use samples of random pairs (random samples, hereafter)
to account for random contamination in candidate binaries. We define the random
sample as a sample of pairs with $20\arcsec<\theta<30\arcsec$ taken from the
initial pool of stellar pairs. Since pairs in the random sample pass the same
data quality selection as candidate binaries, and since virtually all of them
are chance associations ($99.75\%$; see Section~\ref{samples} and
Figure~\ref{ang_sep}), the random sample is a fair representation of the
population of randomly associated stars in candidate binary samples.

\subsection{The Geometric Selection}
\label{geo_select}

\begin{figure}
\epsscale{0.5}
\plotone{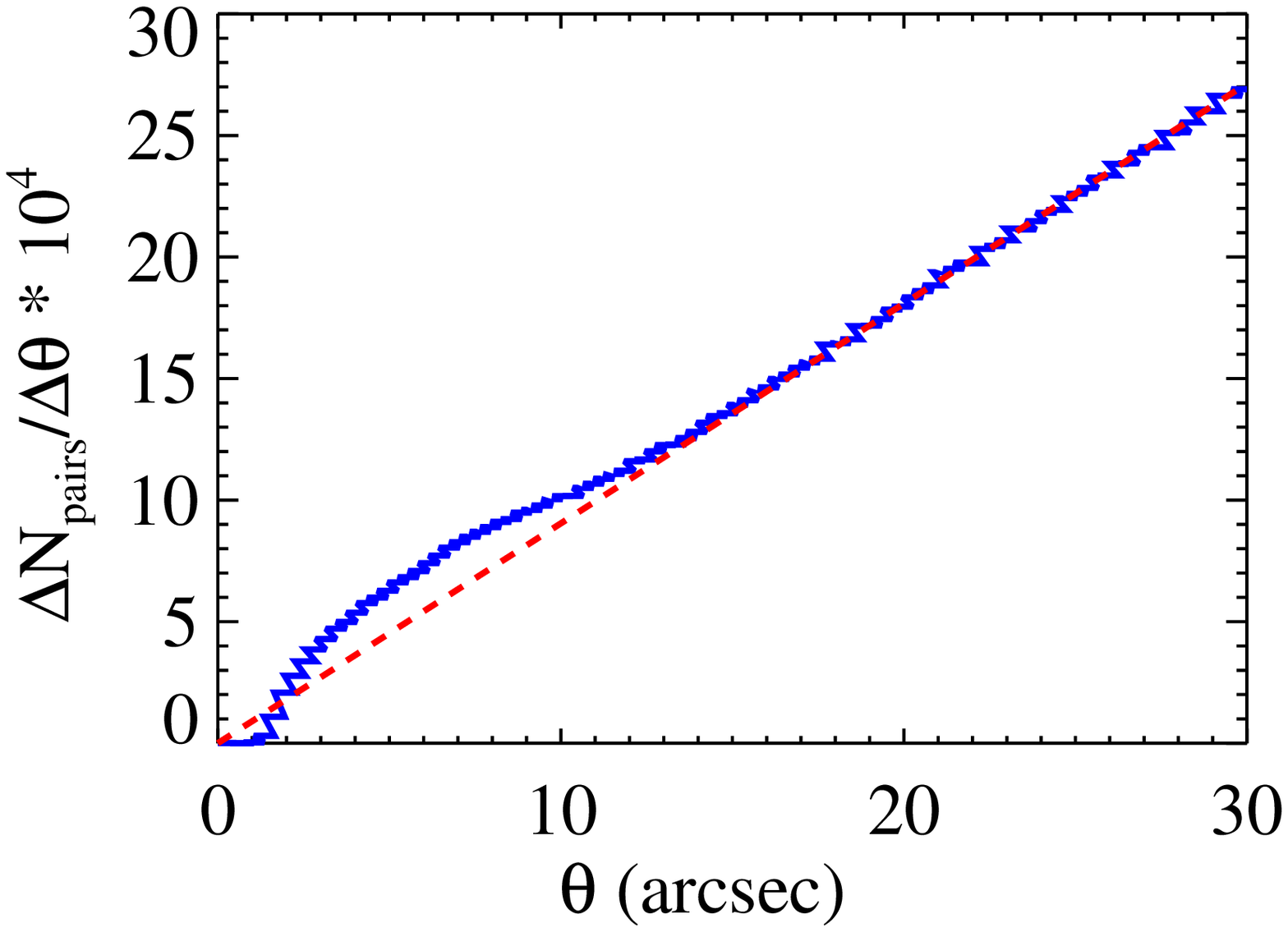}

\plotone{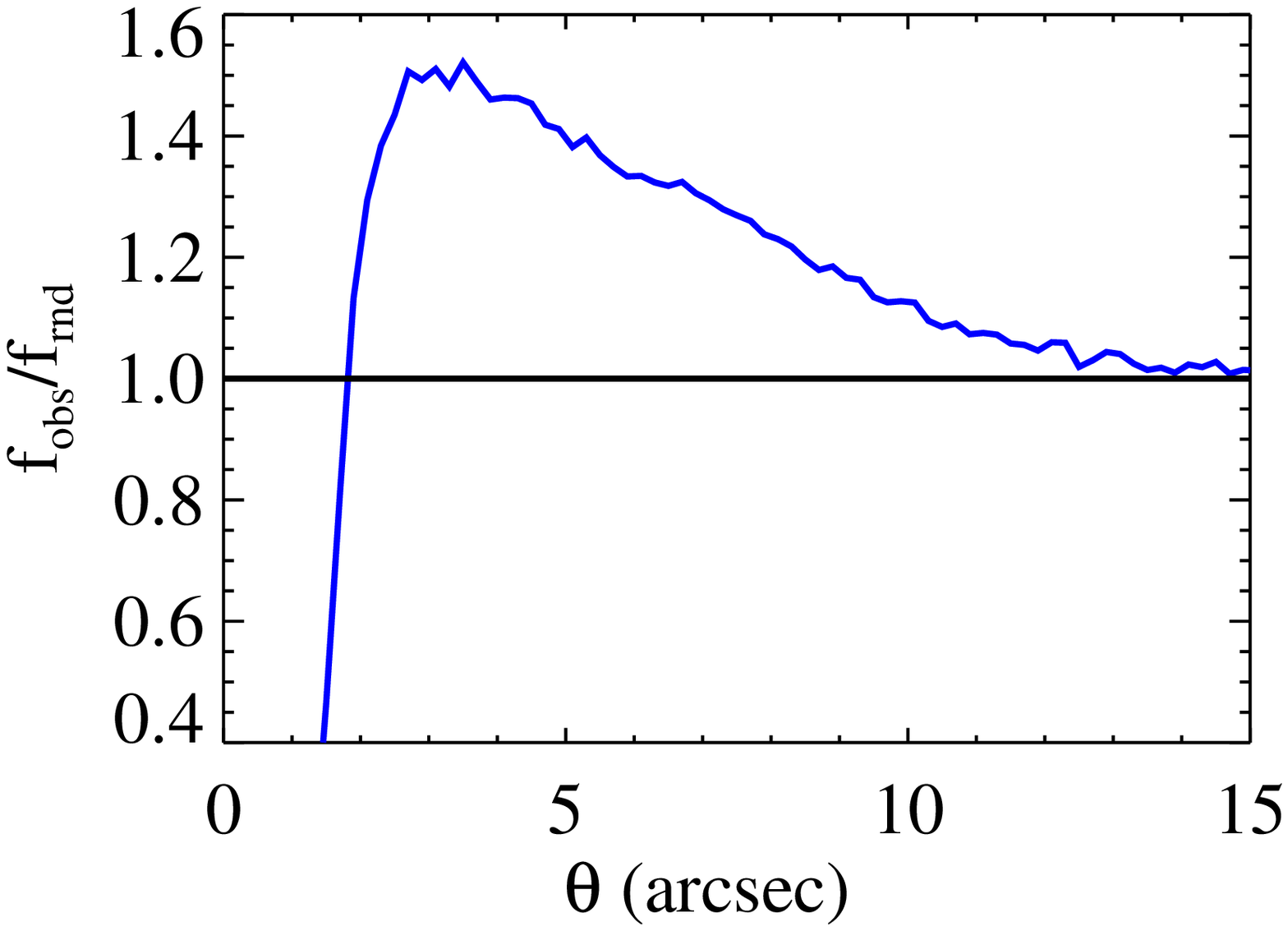}

\plotone{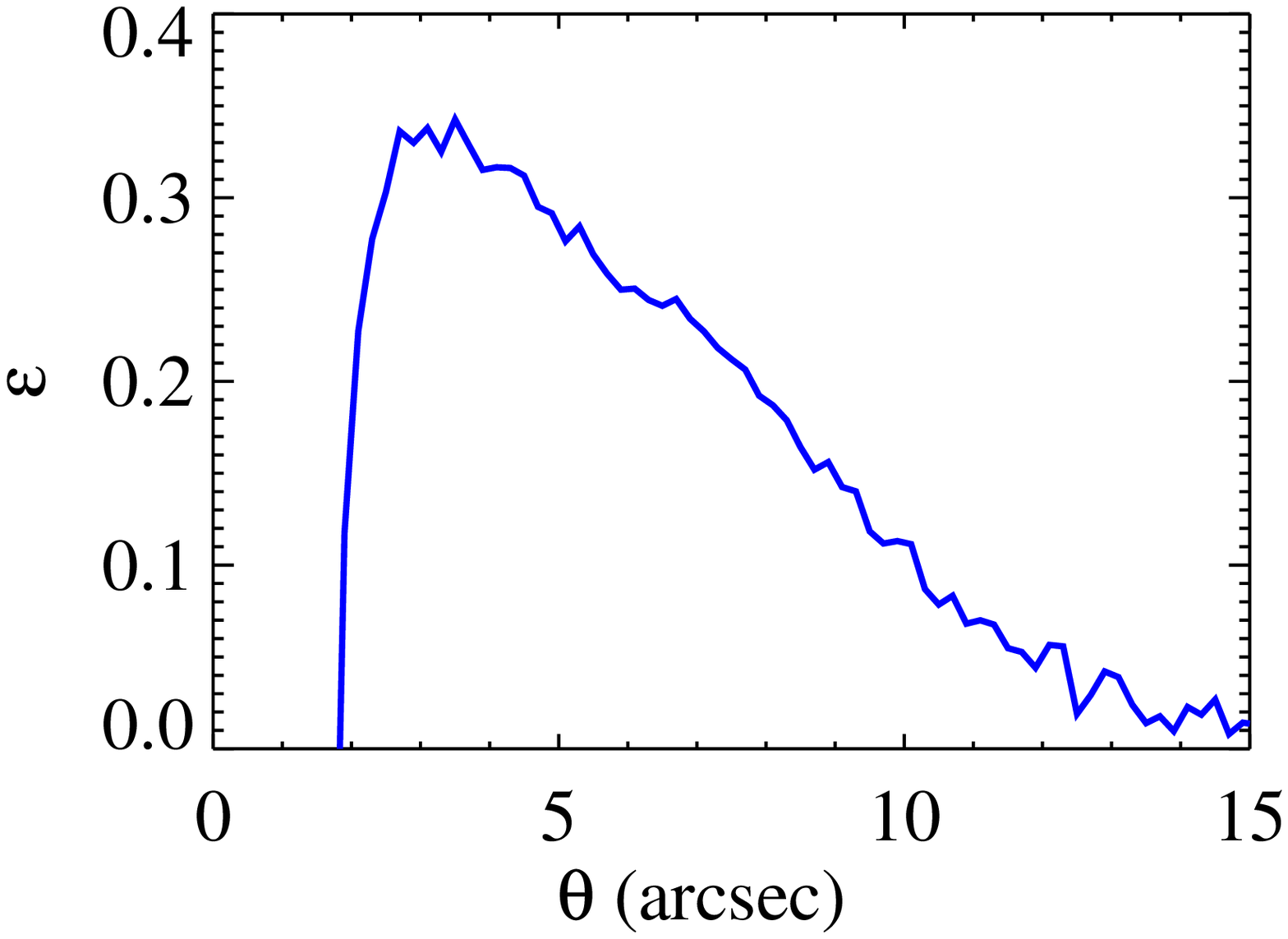}
\caption{
{\em Top:} A comparison of observed ($f_{obs}$, solid histogram) and random
($f_{rnd}$, dashed line, see text) distributions of angular separation $\theta$.
{\em Middle:} Ratio $f_{obs}/f_{rnd}$ as a function of angular separation
$\theta$.
{\em Bottom:} Fraction of true binary systems, $\epsilon$, as a function of
angular separation $\theta$.
\label{ang_sep}}
\end{figure}

The excess of pairs with $\theta<15\arcsec$ in Figure~\ref{ang_sep}
({\em top}) likely indicates a presence of true binaries, and the angular
separation provides a simple, {\em geometric} criterion to select candidate
binary systems. This excess, shown as the ratio $f_{obs}/f_{rnd}$ in
Figure~\ref{ang_sep} ({\em bottom}), increases for $\theta<15\arcsec$, reaches a
relatively flat peak of $\sim1.45$ for $3\arcsec<\theta<4\arcsec$, and sharply
decreases for $\theta<2\arcsec$ due to finite seeing and inability to resolve
close pairs of sources. This excess is related to the fraction of true binaries,
$\epsilon(\theta)$, as
\begin{equation}
\epsilon(\theta) = 1 - f_{rnd}(\theta)/f_{obs}(\theta)\label{eff2}.
\end{equation}
Using Figure~\ref{ang_sep} ({\em bottom}), we choose $3\arcsec<\theta<4\arcsec$
for our geometric selection criterion, since the fraction of true binaries is
expected to reach a maximum of $\sim35\%$ in this range.

The interpretation of the excess of close stellar pairs as gravitationally bound
binary pairs implies that the components are at similar distances. If this is
true, and if it is possible to constrain the distance via a photometric parallax
relation, than their distribution in the color-magnitude diagram should be
different than for a sample of randomly associated stars.

\begin{figure}
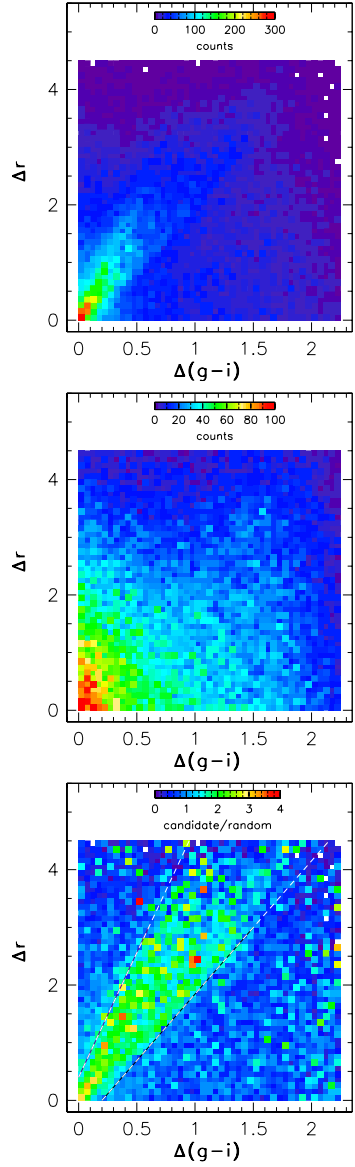

\epsscale{0.3}
\plotone{f2a.eps}

\plotone{f2b.eps}

\plotone{f2c.eps}
\caption{
Distribution of counts for the geometrically-selected candidate sample
({\em top}), random sample ({\em middle}), and the ratio of two maps
({\em bottom}) in the $\Delta r=r_2-r_1$ vs.~$\Delta(g-i)=(g-i)_2 - (g-i)_1$
diagram, binned in $0.05\times0.1$ mag bins. The average candidate-to-random
ratio in the region outlined by the dashed lines (Eq.~\ref{color_cut1}
and~\ref{color_cut2}) is $\sim1.7$, implying that $>40\%$ of candidates are true
binaries.
\label{counts}}
\end{figure}

To test this hypothesis, we select 51,753 candidate binaries with
$3\arcsec<\theta<4\arcsec$. We compare their distribution in the
$\Delta r=r_2-r_1$ vs.~$\Delta(g-i)=(g-i)_2 - (g-i)_1$ diagram to the
distribution of pairs from the random sample, as shown in Figure~\ref{counts}.
The number of pairs in this random sample is restricted to 51,753. Were the
selection a random process, the selected candidates would have the same
distribution in this diagram as the random sample, and the average
candidate-to-random ratio would be $\sim1$. However, in the region where
\begin{equation}
4.33\Delta(g-i) - \Delta r + 0.4 > 0\label{color_cut1},
\end{equation}
and
\begin{equation}
2.31\Delta(g-i) - \Delta r - 0.46 < 0\label{color_cut2}
\end{equation}
the two distributions are different (average candidate-to-random ratio of
$\sim1.7$), implying that $>40\%$ of candidates are found at similar
distances. In principle, a selection cut using Equations~\ref{color_cut1}
and~\ref{color_cut2} could be made to increase the fraction of true binaries in
the candidate sample. We do not make such a cut a priori, but instead develop a
method (described in Section~\ref{method}) that robustly ``ignores'' random
pairs while estimating the photometric parallax relation. After a best-fit
photometric parallax relation is obtained, the contamination can be minimized by
selecting only pairs where both components are at similar distances, as
described in Section~\ref{samples}.

\begin{figure}
\epsscale{0.4}
\plotone{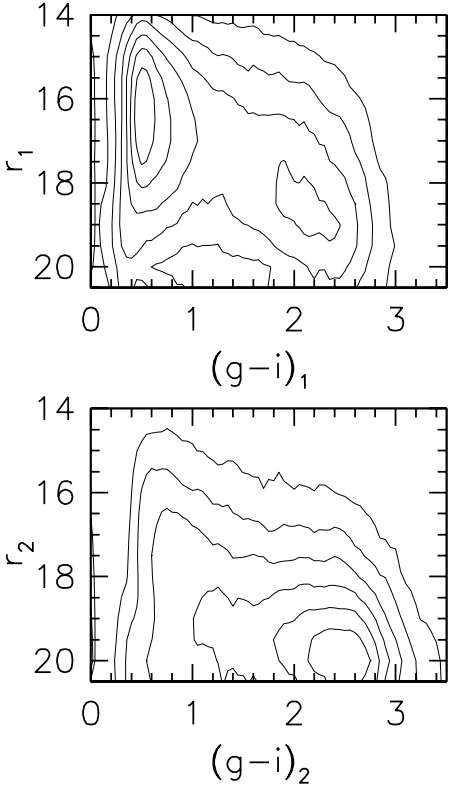}
\caption{
The $r$ vs.~$g-i$ distribution of brighter ({\em top}) and fainter
({\em bottom}) components from the geometrically-selected sample of candidate
binaries, shown with linearly spaced contours.
\label{r_vs_gi}}
\end{figure}

The $r$ vs.~$g-i$ distributions of brighter and fainter components of candidate
binaries are shown in Figure~\ref{r_vs_gi}. We find that the brighter components
in the candidate sample are mostly disk G to M dwarfs, while the fainter
components are mostly M dwarfs.

\subsection{The Kinematic Selection}
\label{kin_select}

As seen from Figure~\ref{ang_sep} ({\em top}), candidate binaries with
$\theta>15\arcsec$ cannot be efficiently selected using angular distance only,
as nearly all pairs in this range are most likely chance associations. In this
regime, a {\em kinematic} selection based on common proper motion should be more
efficient, as random pairs have a small probability ($\sim0.005$ determined
using Monte Carlo simulations) to be common proper motion pairs (using selection
criteria listed below).
 
We therefore select a second sample of 14,148 candidate binaries by searching
for common proper motion pairs with proper motion difference
$\Delta \mu=|\mathbf{\mu_2}-\mathbf{\mu_1}|<5$ mas $yr^{-1}$, and with absolute
proper motion in the range 15 mas
$yr^{-1}<|\mathbf{\mu}|_{max}<$ 400 mas $yr^{-1}$, where
$|\mathbf{\mu}|_{max}=max(|\mathbf{\mu_1}|,|\mathbf{\mu_2}|)$. These criteria
require that the directions of two proper motion vectors agree at a $1\sigma$
level, and that the proper motion is detected at a $5\sigma$ level or higher.
The common proper motion pairs with orbital motion $\ga1\arcsec$ over 50 years
are not selected because their USNO-B and SDSS positions place them outside the
$1\arcsec$ search radius used by Munn et al. The angular separation of common
proper motion pairs is limited to $9\arcsec<\theta<30\arcsec$. Pairs of sources
with $\theta<9\arcsec$ are usually blended in the USNO-B data and may not have
reliable proper motion measurements (see Section~\ref{limitations}), while the
maximum angular separation between sources in the CAS ``Neighbors'' table
defines the upper limit of $\theta<30\arcsec$. However, for purposes of
Section~\ref{limitations}, we have created a sample of common proper motion
pairs that extends to $\theta=500\arcsec$. We have done so by matching
SDSS sources (that pass the quality flags from Section~\ref{initial_select})
within a $500\arcsec$ search radius into common proper motion pairs. Since this
matching is computationally expensive, we have done this only for one sample.
The $r$ vs.~$g-i$ distributions of brighter and fainter components of
kinematically-selected candidate binaries are similar to those shown in
Figure~\ref{r_vs_gi}.

\subsection{The Sample Completeness}
\label{composition}

Before proceeding with the determination of photometric parallax relations and
discussion of the properties of wide binary systems, we summarize the
completeness of geometric and kinematic samples, and estimate their expected
fractions of disk and halo populations. The samples are selected from a
highly-dimensional space of measured parameters and an understanding of the
selection effects is a prerequisite for determining the limitations of various
derived statistical properties. For example, the geometric sample is selected
using five parameters: the $g-i$ color of the two components, $(g-i)_1$ and
$(g-i)_2$, their apparent magnitudes, $r_1$ and $r_2$, and their angular
separation on the sky, $\theta$. The latter three can be transformed with the
aid of a photometric parallax relation into a difference of their apparent
magnitudes, $\Delta m =r_2-r_1$, distance $D$, and the projected physical
separation, $a$. We seek to constrain the photometric parallax relation by
minimizing the difference $\delta = \Delta M - \Delta m$, where $\Delta M$ is a
two-dimensional function of $(g-i)_1$ and $(g-i)_2$ (Section~\ref{method}), and
at the same time derive constraints on the two-dimensional color distribution of
wide binaries, on their $a$ distribution, and on any variation of these
distributions with position in the Galaxy (Section~\ref{properties}). Not all of
these constraints can be derived independently of each other, and most are
subject to severe selection effects. By judiciously selecting data subsets and
projections of this five-dimensional parameter space, these effects can be
understood and controled, as described below. 

\begin{figure}
\epsscale{0.55}
\plotone{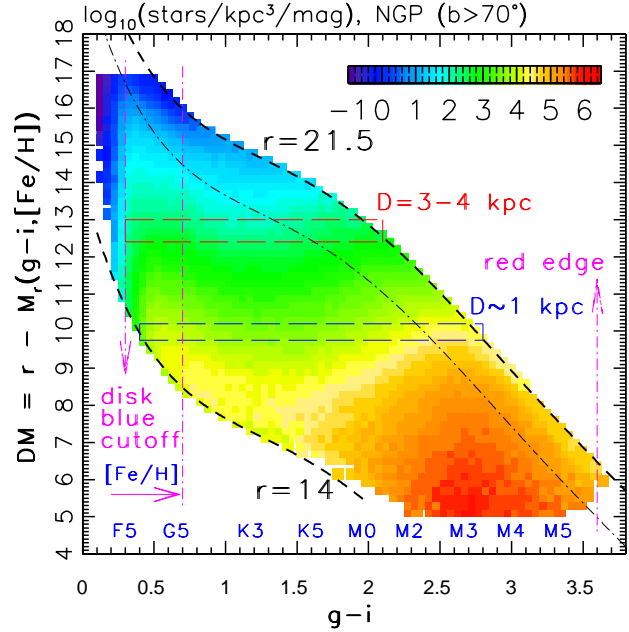}
\caption{The color-coded map, with the legend shown in the top right corner,
shows the logarithm of the volume number density (stars/kpc$^3$/mag) of
$\sim$2.8 million stars with $14<r<21.5$ observed towards the north Galactic
pole ($b>70\arcdeg$), as a function of their distance modulus and the $g-i$
color (the density variation in the horizontal direction represents luminosity
function, and the variation in the vertical direction reflects the spatial
volume density profiles of disk and halo stars). The absolute magnitudes are
computed using expressions A3 and A7 from I08a, and the displayed distance range
is 100 pc to 25 kpc. Stars are color-selected from the main stellar locus
(dominated by main-sequence stars) using criteria 3-5 from Section 2.3.1 in
I08a. The metallicity correction is applied using photometric metallicity for
stars with $g-i<0.7$ (based on Eq.~4 from I08a), and by assuming $[Fe/H]=-0.6$
for redder stars. As illustrated above the $g-i$ axis using the MK spectral type
vs.~$g-i$ color table from \citet{cov07}, this color roughly corresponds to G5.
The two vertical arrows mark the turn-off color for disk stars, and the red edge
of M dwarf color distribution (there are redder M dwarfs detected by SDSS, but 
their volume number density, i.e., the luminosity function, falls precipitously 
beyond this limit; J. Bochanski, priv. comm.). The two diagonal dashed lines 
show the apparent magnitude limits, $r=14$ and $r=21.5$. The dot-dashed
diagonal line corresponds to $r=20$, which approximately describes the 50\% 
completeness limit for stars with cataloged proper motions \citep{mun04}. Around
the marked distance range of 3-4 kpc, the counts of halo stars begin to dominate
disk stars (see Fig.~6 in I08a), and the distance range around 1 kpc offers the
largest color completeness.
\label{Fig:DMvsColor}}
\end{figure}

To illustrate the most important selection effects, we employ the photometric
parallax relation and its dependence on metallicity derived by
\citet[hereafter I08a]{ive08a}. The quantitative differences between their
photometric parallax relation and the ones derived here have negligible impact
on the conclusions derived in this Section. For simplicity, we select a sample
of $\sim$2.8 million stars with $r<21.5$ observed towards the north Galactic
pole ($b>70\arcdeg$), and study their counts as a function of distance and the
$g-i$ color. Due to this choice of field position, the distance to each star is
approximately equal to its distance from the Galactic plane (for a detailed
study of the dependence of stellar number density on position within the Milky
Way, see J08). Figure~\ref{Fig:DMvsColor} illustrates several important
selection effects. 

First, for any $g-i$ color there is a minimum and maximum distance corresponding
to the SDSS saturation limit at $r\sim14$ and the adopted faint limit at
$r=21.5$; the probed distance range extends from 100 pc to 25 kpc. Within the
distance limits appropriate for a given color, the sample is essentially
complete ($\sim$98\%, \citealt{fin00}). Second, these limits are strongly
dependent on color: the bluest stars saturate at a distance of about 1 kpc,
while the reddest stars are too faint to be detected even at a few hundred pc.
Equivalently, due to the finite dynamic range of SDSS apparent magnitudes, there
is no distance range where the entire color range from the blue disk turn-off
edge to the red edge of luminosity function is completely covered. At best, at
distances of about 1 kpc the color completeness extends from the blue edge to
the peak of luminosity function at $g-i \sim 2.7$. Third, when pairing stars
into candidate binary systems, their color distribution at a given distance (the
requirement that the differences of apparent and absolute magnitudes are similar
places the two stars from a candidate pair into a narrow horizontal strip in the
distance modulus (DM) vs.~$g-i$ diagram shown in Figure~\ref{Fig:DMvsColor})
will be clipped: the ratio of the number of candidate binaries and the number of
all single stars in the sample decreases at distances significantly different
from $\sim1$ kpc because of a bias against blue-red pairs.

The binary samples selected from the $\sim$1 kpc distance range can be used to
measure the two-dimensional color distribution of wide binaries, as well as to
gauge the dependence of their $a$ distribution on color. The dependence of the
$a$ distribution on distance from the Galactic plane can also be studied over a
substantial distance range, but {\it only under the assumption that it is
independent of color.} 

The imposed $\theta$ range ($3\arcsec$ to $30\arcsec$) limits the range of
probed physical separation to values proportional to distance, and ranging from
3,000 AU to 30,000 AU at a distance of 1 kpc. We discuss and account for these
effects in more detail in Section~\ref{spatial}.

\subsection{The separation of disk and halo populations}
\label{separation}

The counts of main-sequence stars shown in Figure~\ref{Fig:DMvsColor} include
both disk and halo populations. With the available data, there are three methods
that might be used for separating stars (including candidate binary systems)
into disk and halo populations (Juri\'{c} et al., in prep.):
\begin{enumerate}
\item
A statistical method based on the stellar number density profiles (J08): beyond
about 3 kpc from the plane, halo stars begin to dominate. However, as shown in
Figure~\ref{Fig:DMvsColor}, only stars bluer than $g-i=2$ are detected at such
distances. The stellar number density profiles suggest that the fraction of halo
stars is below $\sim$20\% closer than 1.5 kpc from the Galactic plane (see
Figure~6 in I08a).
\item
Classification based on metallicity into low-metallicity
($\left[Fe/H\right]<-1$) halo stars and higher metallicity stars. As shown by
I08a, this is a robust and accurate method even when using photometric
metallicity estimator, but it works only for stars with $g-i \la 0.7$ due to the
limitations of the photometric metallicity method, and the SDSS spectroscopic
metallicity is available only for a small fraction of stars in the candidate
samples.
\item 
Kinematic selection based on proper motion measurements, and implemented via a
reduced proper motion diagram (e.g., \citealt{sg03}; \citealt{mun04}, and
references therein). However, as discussed in detail in Appendix B, this method
is robust only closer than 2-3 kpc from the Galactic plane due to a rotational
velocity gradient of disk stars which diminishes kinematic differences between
halo and disk stars further away from the plane. 
\end{enumerate}

Given the limitations of these methods, it is not possible to reliably separate
disk and halo populations throughout the explored parameter space, and in both
geometric and kinematic samples. For geometric sample, the third method is not
applicable because SDSS-POSS proper motions are not reliable at small angular
distances ($\theta\la9\arcsec$; see Section~\ref{limitations}). The requirement
$g-i \la 0.7$ required for the second method results in a subsample with too
narrow a color range to constrain the photometric parallax relation.
Nevertheless, the analysis of this subsample based on results from I08a
indicates that fewer than 10\% of stars in geometric sample belong to halo
population (this fraction increases with the distance from the Galactic plane;
see Figure 6 in I08a), and thus we expect that halo contamination plays only a
minor role in the geometric sample. 

The kinematic sample is expected to include a non-negligible fraction of halo
stars due to the selection of stars with substantial proper motions. We use the
reduced proper motion diagram to estimate the fraction of halo candidate binary
stars in this sample. The reduced proper motion for an arbitrary photometric
bandpass, here $r$, is defined as
\begin{equation}
\label{Eq:rpm}
                   r_{RPM} =  r + 5\log{(\mu)},
\end{equation}
where $\mu$ is proper motion in arcsec $yr^{-1}$ (sometimes an additional offset
of 5 mag is added). Using a relationship between proper motion, distance and
tangential velocity,
\begin{equation}
                  v_t = 4.47 \, \mu \, D 
\end{equation}
and 
\begin{equation} 
                  r-M_r = 5\log{(D)}-5, 
\end{equation}
Equation~\ref{Eq:rpm} can be rewritten as 
\begin{equation}
\label{Eq:rpm2}
               r_{RPM} =  M_r + 5\log{(v_t)} + C,
\end{equation}
where $D$ is distance in parsec, $M_r$ is the absolute magnitude, and $v_t$ is
the heliocentric tangential velocity (the projection of the heliocentric
velocity on the plane of the sky), and $C$ is a constant ($C=-8.25$ if $v_t$ is
expressed in km $s^{-1}$). Therefore, for a population of stars with the same
$v_t$, the reduced proper motion is a measure of their absolute magnitude. As
shown using similar data as discussed here, halo and disk stars form two
well-defined and separated  sequences in the reduced proper motion vs.~color
diagram (e.g., \citealt{sg03}; \citealt{mun04}; and references therein). We
discuss the impact of different metallicity and velocity distributions of halo
and disk stars on their reduced proper motion distributions in more detail in
Appendix B.

Figure~\ref{Fig:1wd} shows reduced proper motion diagrams for stars observed 
towards the north Galactic pole, constructed for two ranges of observed proper 
motion: 15-50 mas $yr^{-1}$ and 50-400 mas $yr^{-1}$. The choice of the proper
motion range, together with unavoidable apparent magnitude limits, strongly
affects the probed distance range: the larger is the proper motion, the closer
is the distance range over which the selection fraction is non-negligible. We
find that the two sequences closely follow the expectations based on the
analysis of metallicity and velocity distributions from I08a. The halo sequence
can be efficiently separated by selecting stars with reduced proper motion
larger than a boundary generated using the photometric parallax relation from
I08a, evaluated for the median halo metallicity ($\left[Fe/H\right]=-1.5$) and
with $v_t$=180 km $s^{-1}$ (see Equation~\ref{Eq:rpm2}). This separation method
is conceptually identical to the $\eta$ separator discussed by \citet{sg03}.
They also proposed to account for a shift of the reduced proper motion sequences
with galactic latitude, an effect which we discuss in more detail in Appendix B.
For the reasons described there, to account for the variation of the reduced
proper motion sequences away from the Galactic pole, we simply offset the $v_t$
value from 180 km $s^{-1}$ to 110 km $s^{-1}$ (i.e., the separator moves upwards
in Figure~\ref{Fig:1wd} by 1 mag). While this selection removes some disk
binaries, it is designed to exclude most of halo binaries from the sample.

With the aid of reduced proper motion separator, we separate kinematic sample
into candidate halo (1,336 pairs) and disk binaries (10,112 pairs). This
fraction of halo systems is consistent with the above estimate obtained for the
geometric sample. To assess selection effects, we first investigate the sample
of single stars. The top left panel in Figure~\ref{Fig:2wd} shows the fraction
of all the stars shown in Figure~\ref{Fig:1wd} that have proper motion larger
than 15 mas $yr^{-1}$ and $r<19.5$ (the latter limit ensures the SDSS-POSS
proper motion catalog completeness above $\sim$90\%). The selection efficiency
is a strong function of distance, and falls from its maximum of $\sim$95\% for
nearby stars to below 50\% at a distance of about 1 kpc. The candidate disk
stars are detected in significant numbers to $\sim$3 kpc, and halo stars beyond
$\sim$1 kpc. The fraction of selected stars that are classified as halo stars is
below 20\% closer than $\sim$1.5 kpc from the Galactic plane, and becomes
essentially 100\% beyond 3 kpc. 

The kinematic difference between halo and disk stars is blurred at distances
beyond 2-3 kpc (see Appendix B), and the majority of disk stars at such
distances are misidentified as halo stars (the metallicity distribution implies
that disk stars do exist at distances as large as 7 kpc from the Galactic plane,
see Figure~10 in I08a). To demonstrate this effect, we use subsamples of
candidate disk and halo binaries identified using the reduced proper motion
diagram that have $0.2<(g-r)_1<0.4$. For these pairs it is possible to estimate
photometric metallicity (I08a) and use it as an independent population
classifier. Figure~\ref{plot_ug} shows that practically all candidate binaries
with $\left[Fe/H\right]>-1$ further than $\sim$2 kpc from the Galactic plane are
misclassified as halo stars when using reduced proper motion diagram. 

In summary, geometric sample is heavily dominated by disk binaries, with halo
contamination all but negligible closer than about 2 kpc from the plane.
Kinematic sample becomes severely incomplete ($<$50\%) further than $\sim$2 kpc
from the plane, and has a higher fraction of halo binaries than geometric
sample, at a given distance from the plane. However, this halo contamination can
be efficiently removed using the reduced proper motion diagram. Unfortunately,
the number of selected halo binaries is insufficient in number (1,336 in
kinematic and 5,556 in geometric sample), and spans too narrow a color range to
robustly constrain the photometric parallax relation. Therefore, both samples of
candidate binaries are supposed to yield similar photometric parallax relations,
because both are dominated by disk stars.

\section{The Photometric Parallax Estimation Method}
\label{method}

In principle, both the normalization and the shape of the photometric parallax 
relation (i.e., the shape of the main sequence in the Hertzsprung-Russell
diagram) vary as a function of color and metallicity \citep{lcl88, sie02}. Since
our data do not allow a reliable estimate of metallicity over the entire range
of observed colors, we can only estimate the ``mean'' shape of the photometric
parallax relation as a function of color, for all metallicities present in the
sample. Such a mean shape is approximately an average of individual
$\left[Fe/H\right]$-dependent relations, weighted by the sample metallicity
distribution. J08 derived such ``mean'' photometric parallax relations
appropriate at the red end for the nearby, metal-rich stars, and at the blue end
for distant, metal-poor stars. I08a discuss the offset of photometric parallax
relation as a function of metallicity (see their Figure~20), and derived the
metallicity range implied by ``mean'' photometric parallax relations from J08.
The derived metallicity range is consistent with the spatial distribution of
metallicity derived by I08a and the color-magnitude limits of the SDSS survey.

\subsection{The Photometric Parallax Parametrization}
\label{parametrization}

We adopt the J08 polynomial $r-i$ parametrization of the photometric parallax
relation
\begin{equation}
M_r(r-i|\mathbf{p})=A+B(r-i)+C(r-i)^2+D(r-i)^3+E(r-i)^4\label{photo_parallax},
\end{equation}
where $\mathbf{p}=(A,B,C,D,E)$ are the parameters we wish to estimate. To
improve their accuracy, Juri\'c et al.~used a maximum likelihood technique to
estimate the $r-i$ color from the observed $g-r$ and $r-i$ colors. Because of
the brighter flux limit employed here, we use the measured $g-i$ color to derive
a best estimate of the $r-i$ color via a stellar locus relation (J08):
\begin{equation}
g-i = 1.39(1-exp[-4.9(r-i)^3-2.45(r-i)^2-1.68(r-i)-0.050]) + r - i\label{interp}
\end{equation}
The $r-i$ color estimate obtained with this method has several times smaller
noise than the measured $r-i$ color. This is because the observed dynamic range
for the $g-i$ color is much larger than of the $r-i$ color ($\sim3$ mag
vs.~$\sim1$ mag), while their measurement errors are similar.

\subsection{The Parameter Estimation Algorithm}
\label{algorithm}

The goal of parameter estimation algorithm is to determine the photometric
parallax relation, $M_r(r-i|\mathbf{p})$, that minimizes the width of the
distribution of $\delta$ values for {\em true} binary systems, where
$\delta = (M_{r2}-M_{r1})-(r_2-r_1)$. The $\chi^2$ minimization cannot be used
for this purpose because random pairs, if not removed from the sample, will
strongly bias the best-fit $M_r$. The available color, angular separation, and
proper motion information are insufficient to separate the random pairs from the
true binaries. Therefore, we need to design a fitting algorithm that will be
least affected as possible by random pairs.

We begin by studying the behavior of $\delta$ values in mock catalogs. The first
step in creating a mock catalog is the selection of 51,753 (random) pairs from
the random sample. Note that the fraction of true binaries in the random sample
is only $\sim0.25\%$ (see Section~\ref{samples}). True binaries are then
``created'' in the mock catalog by replacing the observed $r_2$ magnitudes for
$20\%$ of pairs with
\begin{equation}
r_2 = r_1 + (M_{r2}-M_{r1}) + N(0,0.1)\label{r2_mag},
\end{equation}
where $M_r=M_r(r-i|\mathbf{p_0})$ and
$\mathbf{p_0}=(3.2,13.30,-11.50,5.40,-0.70)$ (Equation 2 coefficients from J08).
The $N(0,0.1)$ is a Gaussian random variate added to account for the intrinsic
scatter around the photometric parallax relation. The result of this process is
a mock sample of candidates where $20\%$ of pairs are ``true'' binaries, and the
rest ($80\%$) is the contamination made of random pairs. The distribution of
$\delta$ values for ``true'' binaries is, by definition, a 0.1 mag wide Gaussian
centered on zero when $M_r=M_r(r-i|\mathbf{p_0})$.

Figure~\ref{delta_hist_mock} ({\em top}) shows the distribution of $\delta$
values for the mock sample evaluated with the ``true'' [$M_r(r-i|\mathbf{p_0})$]
photometric parallax relation. The observed $\delta$ distribution can be
described as a sum of a Gaussian and a non-Gaussian component. The non-Gaussian
component is due to random pairs (the contamination), while the Gaussian
component (0.1 mag wide and centered on zero) is due to the true binaries.

When an $M_r$ relation different from the ``true'' (or best-fit) $M_r$ is
adopted, the Gaussian component becomes wider and {\em the peak height of the
$\delta$ distribution decreases}, as shown in Figure~\ref{delta_hist_mock}
({\em bottom}). At the same time, the peak height of the $\delta$ distribution
of the contamination changes much less since the distribution is much wider
($\sim2.3$ mag wide). Therefore, {\em minimizing} the width of the $\delta$
distribution of true binaries, is equivalent to {\em maximizing} the peak height
of the entire $\delta$ distribution. We quantify this peak height as the number
of candidate binaries in the most populous $\delta$ bin.

\subsection{The Algorithm Implementation}
\label{implementation}

To robustly explore the parameter space that defines the photometric parallax
relation, and to find the best-fit coefficients $\mathbf{p}$, we implement our
algorithm as a Markov chain Monte Carlo (MCMC) process. The MCMC description
given here and our implementation of the algorithm are based on examples given
by \citet{teg04}, \citet{for05}, and \citet{cro06}.

The basic idea of the MCMC aproach is to take an {\em n}-step intelligent random
walk around the parameter space while recording the point in parameter space for
each step. Each successive step is allowed to be some small distance in
parameter space from the previous position. A step is always accepted if it
improves the fit, and is sometimes accepted on a random basis even if the fit is
worse, where the goodness of the fit is quantified by some parameter (usually
with $\chi^2$). The random acceptance of a bad fit ensures that the MCMC does
not become stuck in a local minimum, and allows the MCMC to fully explore the
surrounding parameter space.

We start a Monte Carlo Markov chain by setting all coefficients from
Equation~\ref{photo_parallax} to zero ($\mathbf{p_i}=0$). Using this initial set
of coefficients we evaluate $\delta=(M_{r2}-M_{r1})-(r_2-r_1)$ for all candidate
binaries assuming $M_r(r-i|\mathbf{p_i})$, and bin $\delta$ values in 0.1 mag
wide bins. The number of candidate binaries in the most populous bin, $P_i$, is
used to quantify the relative goodness of the fit.

Given $\mathbf{p_i}$, a new candidate step,
$\mathbf{p_n}=\mathbf{p_i}+\Delta\mathbf{p}$, is generated, where the step size,
$\Delta \mathbf{p}$, is a vector of independent Gaussian random variates with
initial widths, $\sigma$, set to 1. Using the candidate set of coefficients,
$\mathbf{p_n}$, $\delta$ values are evaluated, binned, and the height of the
$\delta$ distribution is assigned to parameter $P_n$.

Following the Metropolis-Hastings rule \citep{met53,has70} the candidate step is
accepted ($\mathbf{p_{i+1}}=\mathbf{p_n}$, $P_{i+1}=P_n$) if $P_n > P_i$ or if
$exp(P_n - P_i)>\xi$, where $\xi$ is a random number between 0 and 1
($\xi\in[0,1]$). Otherwise, the candidate step is rejected.

While the Metropolis-Hastings rule guarantees that the chain will converge, it
does not specify {\em when} the convergence is achieved. The speed of the
convergence depends on the Gaussian scatter $\sigma$ used to calculate the step
size $\Delta \mathbf{p}$. If the scatter is too large, a large fraction of
candidate steps is rejected, causing the chain to converge very slowly. If the
scatter is too small, the chain behaves like a random walk, and the number of
steps required to traverse some short distance in the parameter space scales as
$1/\sigma^2$. The choice of optimal Gaussian scatter $\sigma$ (for each fitted
coefficient), as a function of the position in the parameter space, is not
trivial and it can be very complicated even if the fitted coefficients are
uncorrelated.

To determine the optimal $\sigma$ values we follow the \citet{teg04}
prescription (see their Appendix A). After every 100 accepted steps we compute
the coefficient covariance matrix
$\mathbf{C} = \langle \mathbf{p} \mathbf{p}^t \rangle - \langle \mathbf{p} \rangle \langle \mathbf{p}^t \rangle$
from the chain itself, diagonalize it as
$\mathbf{C}=\mathbf{R} \Lambda \mathbf{R}^t$, and use it to calculate a new step
size $\Delta \mathbf{p'}=\mathbf{R}^t \mathbf{\Lambda}^{1/2} \Delta \mathbf{p}$
for each coefficient separately. We find that this transformation greatly
accelerates the convergence of a chain.

Due to the stochastic nature of the MCMC, the best-fit relations (coefficients
with the highest $P_i$ value in a chain) from different chains will not
necessarily be the same. To quantify the intrinsic scatter between different
best-fit relations, we run fifty 10,000-element long chains, and select the
best-fit coefficients from each chain for subsequent comparison (see
Section~\ref{robustness}). The proper mixing and convergence of chains is
confirmed using the Gelman \& Rubin $R$ statistic \citep{gr92}. Gelman \& Rubin
suggest running the chains until $R<1.2$ for all fitted coefficients. With
10,000 elements in each chain, we obtain $R<1.01$ for all fitted coefficients.

In the end, we select $\mathbf{p}=(A,B,C,D,E)$ with the highest $P_i$ value
among all chains as our best-fit relation. The constant term $A$ is not
constrained with our algorithm, because $A$ (from $M_{r2}$ and $M_{r1}$) cancel
out when evaluating $\delta$. Instead, we constrain $A$ by requiring $M_r=10.07$
at $r-i=1.1$, obtained from trigonometric parallaxes of nearby M dwarfs
\citep{wwh05}.

\subsection{Algorithm Robustness Test}
\label{robustness}

To test the robustness of our algorithm, we apply it to the mock sample
described in Section~\ref{algorithm}. The best-fit relations (obtained from
Markov chains) are compared on a $0.1\leqslant r-i\leqslant1.5$ grid in 0.01 mag
steps. We find an rms scatter of 0.05 mag between Markov chains, and 0.05 mag
rms scatter between the true and the best-fit relation with the highest $P_i$
value.

We repeat this test with a mock sample containing $30\%$ of true binaries. The
rms scatter between the best-fit relations decreases to 0.03 mag, and the rms
scatter between the true and the best-fit relation with the highest $P_i$ value
decreases to 0.01 mag.

Even when only $20\%$ of sources are true binaries (i.e., contamination by
random pairs is $80\%$) our algorithm recovers the ``true'' photometric parallax
relation at the 0.05 mag (rms) level. The accuracy of the fit increases (to 0.01
mag rms) as the contamination decreases (from $80\%$ to $70\%$).

\subsection{Best-fit Photometric Parallax Relations}\label{analysis}

We apply the method described in Section~\ref{implementation} to two samples of
candidate binaries and obtain the best-fit photometric parallax relations
\begin{equation}
M_r = 3.32 + 15.02(r-i) - 18.58(r-i)^2 + 13.28(r-i)^3 - 3.39(r-i)^4
\label{Mr_geo}
\end{equation}
\begin{equation}
M_r = 3.42 + 13.75(r-i) - 15.50(r-i)^2 + 10.40(r-i)^3 - 2.43(r-i)^4
\label{Mr_kin}
\end{equation}
for the geometrically- and kinematically-selected samples, respectively.
Candidate halo binaries were removed from the kinematically-selected sample
using reduced proper motion diagrams (Section~\ref{separation}) before the
Equation~\ref{Mr_kin} was derived. The photometric parallax relations for halo
stars cannot be robustly constrained using geometrically- or
kinematically-selected halo binaries because the color range they span is too
narrow ($g-i < 1.0$ at 3-4 kpc, see Figures~\ref{Fig:DMvsColor}
and~\ref{Fig:2wd}).

We test the correctness of the shape by studying the dependence of median
$\delta$ values on the $g-i$ colors of the brighter and the fainter components.
If the {\em shape} of these photometric parallax relations is correct, the
distribution of $\delta$ values will be centered on zero, and the individual
$\delta$ values will not correlate with color. The medians are used because they
are more robust to outliers (random pairs in the sample). We start by
calculating $\delta$ values for each candidate binary sample (using the
appropriate $M_r$ relation), and then select candidates with $|\delta|<0.4$.
This cut reduces the contamination by random pairs, as demonstrated in
Section~\ref{scatter}. The selected candidate binaries are binned in $g-i$
colors of the brighter and the fainter component, and the median $\delta$ values
are shown in Figure~\ref{medians}.

The distributions of the median $\delta$ for each pixel are fairly narrow
(0.07 mag), and centered on zero. Irrespective of color and the choice of the
two best-fit photometric parallax relations, the deviations are confined to the
0.25 mag range, placing an upper limit on the errors in the mean shape of the
adopted relations.

In Figure~\ref{Mr_ri} we compare the adopted photometric parallax relations 
to J08 ``faint''
\begin{equation}
M_r = 4.0 + 11.86(r-i) - 10.74(r-i)^2 + 5.99(r-i)^3 - 1.20(r-i)^4
\label{Mr_faint}
\end{equation}
and ``bright''
\begin{equation}
M_r = 3.2 + 13.30(r-i) - 11.50(r-i)^2 + 5.40(r-i)^3 - 0.70(r-i)^4
\label{Mr_bright}
\end{equation}
photometric parallax relations. The rms difference between
Equations~\ref{Mr_geo} and~\ref{Mr_kin}, and Equation~\ref{Mr_bright} is
$\sim0.13$ mag, comparable to the rms difference between our
Equations~\ref{Mr_geo} and~\ref{Mr_kin} ($\sim0.13$ mag). The maximum difference
between Equations~\ref{Mr_geo} and~\ref{Mr_kin}, and Equation~\ref{Mr_bright} is
$\sim0.25$ mag, again comparable to the maximum difference between our
Equations~\ref{Mr_geo} and~\ref{Mr_kin} ($\sim0.25$ mag). The different color
distributions of the two samples, shown in Figure~\ref{geo_kin_gi1_gi2_comp},
together with metallicity effects, is the most likely explanation for
differences between the two photometric parallax relations.

\subsection{The Analysis of the Scatter in Predicted Absolute Magnitudes}
\label{scatter}

The scatter in $\delta$ values can be expressed as
\begin{equation}
\langle \delta^2 \rangle = \langle (\Delta M - \Delta m)^2 \rangle \approx \langle \Delta M^2 \rangle + \langle \Delta r^2 \rangle \label{delta_scatter},
\end{equation}
where $\langle \Delta M^2 \rangle$ is the scatter in predicted absolute
magnitudes, and $\langle \Delta m^2 \rangle$ is the scatter in measured apparent
magnitudes. Since the photometric uncertainties of SDSS are well understood, the
intrinsic scatter around the $M_r(r-i)$ relation is possible to measure and
characterize.

In Figure~\ref{delta_hists} we plot the observed distributions of $\delta$
values for the geometrically- and kinematically-selected binaries, and overplot
the $\delta$ distribution of the random sample. The $\delta$ values for the
random sample were calculated with Equations~\ref{Mr_geo} and~\ref{Mr_kin},
respectively. The $\delta$ distribution of the random sample was fitted to the
observed $\delta$ distribution in the $|\delta|>1$ range using the
Kolmogorov-Smirnov test.

By comparing the random and the observed $\delta$ distributions, we find that
the two match well for $|\delta|>1$ (the Kolmogorov-Smirnov test reports
$P\sim0.95$), indicating that candidate binaries with $|\delta|>1$ are almost
certainly random pairs. On the other hand, as $\delta$ approaches zero, the two
distributions become remarkably different ($P\sim10^{-7}$ for $|\delta|<1$),
indicating that these candidate binaries are dominated by true binary systems,
and not by random pairs.

The $\delta$ distribution for true binaries (Figure~\ref{delta_hists}, dashed
line), obtained by subtracting the random from the observed $\delta$
distribution, is clearly not Gaussian. It can be modeled as a sum of two
Gaussian distributions (``narrow'' and ``wide'') centered close to zero, and
about 0.1 mag and 0.55 mag wide. The centers, widths, and areas for the best-fit
Gaussian distributions are given in Table~\ref{gauss}.

To determine the consistency of the observed scatter with photometric errors, we
normalize the $\delta$ values for the kinematically-selected sample with
expected formal errors,
\begin{equation}
\sigma_{\delta} = (\sigma_{M_{r2}}^2 + \sigma_{M_{r1}}^2 + \sigma_{r_2}^2 + \sigma_{r_1}^2)^{1/2},
\end{equation}
and plot the $\delta/\sigma_{\delta}$ distribution in Figure~\ref{delta_norm}.
The $\delta/\sigma_{\delta}$ distribution for true binaries is not a Gaussian
with a width of 1, as we would expect if the scatter in the $\delta$
distribution was only due to photometric errors in the $gri$ bands (note that
the expected random error in $M_r$ is about 5-10 times larger than the random
error of the $g-i$ color because $dM_r/d(g-i)$ varies from $\sim$10 at the blue
edge to $\sim$5 at the red edge).

The width of $\delta/\sigma_{\delta}$ distribution for the
geometrically-selected candidate binaries is about 3 times smaller than in the
kinematically-selected sample. We find that this is due to {\em overestimated}
photometric errors in the geometrically-selected sample, as shown in
Figure~\ref{photo_errors}. The overestimated photometric errors in the $gri$
bands overestimate the expected formal error $\sigma_{\delta}$, and the overall
$\delta/\sigma_{\delta}$ distribution is too narrow. We speculate that the small
angular separation ($\sim3\arcsec$) between the components is the cause of
overestimated photometric errors (perhaps due to sky background estimates). The
small angular separation of components does not affect the magnitudes of stars
in the geometrically-selected sample. If it did, the two $\delta$ distributions
would be significantly different which, as shown in Figure~\ref{delta_hists}, is
not the case.

The observed non-Gaussian scatter in predicted absolute magnitudes may be due to
photometric parallax variation as a function of metallicity. As noted at the
beginning of Section~\ref{method}, we can only estimate the ``mean'' shape of
the photometric parallax relation. Since the intrinsic photometric parallax for
a given wide binary system is different from the mean relation, $\Delta M$ (the
difference of predicted absolute magnitudes) and $\Delta m$ (the measured
difference of apparent magnitudes) will differ. This discrepancy will increase
for systems where the components have significantly different colors.

To test the assumption that the shape of photometric parallax relation increases
the scatter in predicted absolute magnitudes, we use the mock sample constructed
in Section~\ref{algorithm} and add a color-dependent offset to apparent
magnitudes
\begin{eqnarray}
r'_1 = r_1 + \xi (g-i)_1 \\
r'_2 = r_2 + \xi (g-i)_2
\end{eqnarray}
where $\xi$ is a random number between zero and one (the same for both
components). These color-dependent offsets simulate the change in the shape of
the photometric parallax relation due to metallicity. We apply the algorithm
described in Section~\ref{implementation} to the mock sample, and obtain a
revised photometric parallax relation. Using this relation, we analyze the
distribution of $\delta$ values and find that it can be modeled as a sum of two
Gaussians centered on zero, with widths of 0.1 and 0.3 mag. This result suggests
that the non-Gaussian scatter observed in candidate samples may be caused by
the difference between the shapes of the mean photometric parallax relation
and a true relation for a given metallicity (and perhaps other effects, such as
age).

This model-based conclusion is consistent with a direct comparison of relations
derived here and the relations from I08a evaluated for the median halo
metallicity ($\left[Fe/H\right]=-1.5$) and the median disk metallicity
($\left[Fe/H\right]=-0.7$ for distances probed by our sample; see Figure~5 in
the above paper). The two relations corresponding to halo and disk stars are
offset by 0.6 mag due to metallicity difference. Our relations match the
low-metallicity relation at the blue end and the high-metallicity relation at
the red end. Therefore, in the worst case scenario of extremely blue ($r-i=0.3$)
and red ($r-i=1.4$) disk stars, the maximum error in the difference of their
absolute magnitudes is 0.6 mag. When convolved with the observed color
distribution of pairs, the expected scatter is about 0.2-0.3 mag, consistent
with the observed and simulated widths of the $\delta$ distributions.

Unresolved binarity of components in candidate samples may also contribute to
the non-Gaussian scatter in predicted absolute magnitudes. The multiplicity
studies of G dwarfs \citep{dm91} and M dwarfs \citep{fm92} find that a
significant fraction of G and M dwarf stars ($40-60\%$) are unresolved binary
systems. If a component of a wide binary system is an unresolved binary system,
its luminosity will be underestimated (with the magnitude of the offset
depending on the actual composition of the binary) and the $\delta$ value for
the wide binary system will systematically deviate from zero. In
Appendix~\ref{model} we model the presence of unresolved binaries in wide binary
systems, and find that the model can explain the observed $\delta$ scatter.

Therefore, both the intrinsic variations of the photometric parallax relation
and unresolved binaries can explain the observed non-Gaussian scatter of
$\delta$. The data discussed here are insufficient to disentangle these two
effects.

Finally, the uncertainty in predicted absolute magnitudes (error distribution 
for photometric parallax method) can be obtained by drawing random values, $x$,
from a non-Gaussian distribution
\begin{equation}
f(x) = A_1 \, N(x|\mu_1, \sigma_1/\sqrt{2}) + A_2 \, N(x|\mu_2, \sigma_2/\sqrt{2}), 
\label{Mr_scatter}
\end{equation}
where $N(x|\mu,\sigma)$ are Gaussian distributions, and the best-fit parameters
are listed in Table~\ref{gauss}.

\section{The Properties of Wide Binaries}
\label{properties}

The best-fit photometric parallax relation can be utilized to further refine the
samples of candidate binaries and to address questions about their dynamical and
physical properties such as
\begin{itemize}
\item Do wide binaries have the same spatial distribution as single stars?
\item Do wide binaries have the same color distribution as single stars?
\item Are the color distributions of components in wide binary systems 
      consistent with random pairings? 
\item What is the distribution of semi-major axis for wide binaries?
\item Does the distribution of semi-major axis vary with the position in
      the Galaxy?
\end{itemize}

\subsection{High-Efficiency Samples of Candidate Binaries}
\label{samples}

We use the best-fit photometric parallax relations to select samples of
candidate binaries with high selection efficiency (high fraction of true
binaries) by imposing further constraints on $\delta$ values in geometric and
kinematic sample.

As shown in Figure~\ref{delta_hists}, the fraction of random pairs in the
candidate sample is simply $A_{random}/A_{observed}$, where $A_{random}$ and
$A_{observed}$ are the integrals of the random (triangles) and total (thick
solid line) $\delta$ distributions. The fraction of true binaries, or the
{\em selection efficiency}, is then
\begin{equation}
\epsilon = 1 - A_{random}/A_{observed}\label{eff}
\end{equation}
Without a cut on $\delta$, the fraction of true binaries (the selection
efficiency) in the geometrically- and kinematically-selected samples is $34\%$
and $35\%$, respectively. It is reassuring to find that the $\epsilon$ value for
the geometrically-selected sample obtained here, and the one measured in
Section~\ref{geo_select} match so well (at a $1\%$ level), even though the two
methods for estimating $\epsilon$ are independent.

The selection efficiency of $35\%$ for the kinematically-selected sample may
seem low, given that only $0.5\%$ of random pairs pass the common proper motion
criteria. This points to a low fraction of true binaries with angular separation
greater than $15\arcsec$. If this fraction is about 1/400 ($0.25\%$), the common
proper motion criteria will select 2 random pairs (0.5\% out of a 400), and only
1 true binary system. Therefore, $66\%$ of the sample (2 out of 3) will be
random pairs, and $34\%$ (1 out of 3) will be true binary systems, similar to
what we find for the kinematically-selected sample. The result that only 1/400
pairs with $\theta>15\arcsec$ are true binaries puts the fraction of random
pairs in the random sample at $99.75\%$.

Figure~\ref{delta_hists} shows that the true binaries have a much smaller
range of $\delta$ values than the random pairs. Therefore, a cut on $\delta$
would reduce the contamination, and increase the fraction of true binaries in a
sample. By requiring $|\delta|<0.4$, we construct samples where $63\%$ and
$64\%$ of candidates are true binaries. The numbers of candidate binaries in
these cleaner samples are 16,575 (geometrically-selected) and 5,157 candidates
(kinematically-selected), with the expected total number of true binaries about
13,743. The sample efficiency for the geometric sample can be further increased
to 90\% by requiring $|\delta|<0.2$ and $Z < 0.3$ kpc, where $Z$ is the height
above the Galactic plane. Compared to the existing catalogs of wide binaries by
\citet{cg04} and \citet{lb07}, our samples represent a 20-fold increase in the
number of candidate binaries and probe much deeper into the halo (to $\sim4$
kpc). Although a non-negligible fraction of candidate pairs are due to random
pairings ($\sim35\%$), the increase in the number of potential physical pairs is
substantial. 

We emphasize that our method only selects candidates where both components are
{\em main-sequence} stars, while rejecting systems where one of the components
has evolved off the main sequence. This is due to the photometric parallax
relation, as defined here, being correct for main-sequence stars only. Together
with the small expected fraction of giant stars in our sample due to faint
apparent magnitudes (1-2\%, \citealt{fin00}; I08a), this bias results in
practically pure main-sequence sample. We note that the application of a
photometric parallax relation that corresponds to some mean metallicity
distribution introduces systematic errors in estimated $M_r$. We partially
mitigate this problem by averaging distances determined for each binary
component (using Equation~\ref{Mr_geo}). Based on the behavior of photometric
parallax relations and $\delta$ distribution discussed in Section~\ref{scatter},
the systematic uncertainty in obtained distances is most likely not larger than
10-15\% (an understimate due to faint bias for blue stars). Another source of
overall systematic uncertainty in distances is the normalization of
Equation~\ref{Mr_geo} adopted from \citet{wwh05}. This normalization corresponds
to nearby ($<$100 pc) metal-rich stars, while most stars in our sample are
distances of the order 1 kpc. The disk metallicity gradient discussed by I08a
implies systematic distance overestimate of about 10-20\%, partially cancelling
the above underestimate. These systematic uncertainties propagate as systematic
uncertainties of derived semi-major axes discussed in Section~\ref{spatial}. 

\subsection{The Color Distribution of Wide Binaries}
\label{color}

The luminosity of a main-sequence star, and thus its color via photometric
parallax relation, can be used as a proxy for stellar mass. The color-color
distribution of wide binaries, therefore, provides constraints on the
distribution of stellar masses in wide binary systems. To find the color
distribution of wide binaries, we select a volume-complete ($0.7 < d/kpc < 1.0$)
subsample of geometrically-selected candidate binaries with $|\delta|<0.4$, and
plot their distribution in the $(g-i)_2$ vs.~$(g-i)_1$ color-color diagram in
Figure~\ref{counts_0.7_1.0} ({\em top}). The sample is complete in
the $0.4 < g-i < 2.8$ color and 4,200 AU $<a<$ 10,000 AU semi-major axis range
(see Section~\ref{spatial}). Even though the $|\delta|<0.4$ cut increases the
fraction of true binaries, about 14\% of candidates (in the $0.7 < d/kpc < 1.0$
range) are still random pairs that contaminate the map. To remove the
contamination, first we select pairs from the random sample (see the end of
Section~\ref{initial_select}) with $|\delta|<0.4$ and $0.7<d/kpc<1.0$. The
$|\delta|<0.4$ cut on the random sample will not increase the fraction of true
binaries ($\epsilon$) above $\sim1\%$ because $\epsilon$ decreases rapidly with
$\theta$ (see Figure~\ref{logZ_plots} ({\em middle left}) in
Section~\ref{spatial}), and the pairs in the random sample have
$\theta>20\arcsec$. The $(g-i)_2$ vs.~$(g-i)_1$ distribution of this random
sample is shown in Figure~\ref{counts_0.7_1.0} ({\em middle}). The maps are
essentially probability density maps as pixels sum to 1. To correct for the
contamination in the top map, we multiply each pixel in the random map with 0.14
(that being the contamination in the candidate binary sample), and subtract two
maps. The corrected map, presented in Figure~\ref{counts_0.7_1.0}
({\em bottom}), shows that the color-color distribution of true binary systems
is fairly uniform, has a local maximum around $(g-i)_{1,2}\sim2.5$, and reflects
the underlying luminosity function which peaks for red stars
(c.f.~Figure~\ref{Fig:DMvsColor}). 

The map shown in Figure~\ref{counts_0.7_1.0} ({\em bottom}) describes the
probability density, $P[(g-i)_1,(g-i)_2]$, of a wide binary system with
components that have $(g-i)_1$ and $(g-i)_2$ colors falling into a given pixel.
This probability density can be expressed as a product
\begin{equation}
P[(g-i)_1,(g-i)_2] = P[(g-i)_B|(g-i)_A] \, P[(g-i)_A]
\label{eq_wide_binary_prob}
\end{equation}
where $P[(g-i)_B|(g-i)_A]$ is the conditional probability density of having one
component with $(g-i)_B$ color in a wide binary system where the other component
has $(g-i)_A$, and $P[(g-i)_A]$ is the probability density that a star with
$g-i=(g-i)_A$ color is in a wide binary system. These probability densities may
also vary with Galactic coordinates (e.g., with the height above the Galactic
plane), but we cannot study such effects directly because the samples are
volume-complete only in the $0.7 < d/kpc < 1.0$ range.

The conditional probability density, $P[(g-i)_B|(g-i)_A]$, can be extracted from
Figure~\ref{counts_0.7_1.0} ({\em bottom}) map by selecting pixels where either
$(g-i)_1=(g-i)_A$, or $(g-i)_2=(g-i)_A$. The resulting $P[(g-i)_B|(g-i)_A]$ for
several values of $(g-i)_A$ are shown in Figure~\ref{wide_binary_prob}. Red
stars ($(g-i)_A \ga 2.0$) are more likely to be associated with another red star
than with a blue star, while for blue stars the companion color distribution is
flat. The best-fit analytic functions that describe the observed trends are
given in Table~\ref{tbl_wide_bin_prob}.

The probability density, $P[(g-i)_A]$, that a star with $g-i=(g-i)_A$ is in a
wide binary system can be derived by comparing the $g-i$ color distribution of
stars in wide binary systems with the $g-i$ color distribution of all the stars
in the same volume. As shown in Figure~\ref{prob_gi} ({\em top}), the $g-i$
color distribution of stars in the volume-complete wide binary sample roughly
follows the $g-i$ color distribution of all the stars in the same volume. The
ratio of the two distributions (renormalized to an area of 1) gives the
$P[(g-i)_A]$, and is shown in the bottom panel.

The probability for a star to be in a wide binary system ($P[(g-i)_A]$) is
independent of its color. Given this color, the companions of red components
seem to be drawn randomly from the stellar luminosity function, while blue
components have a larger blue-to-red companion ratio than expected from
luminosity function. These results are consistent with recent results by
\citet{lb07}. The overall fraction of stars in wide binary systems is discussed
in the next section.

\subsection{The Spatial Distribution of Wide Binaries}
\label{spatial}

If the semi-major axis distribution function, $f(a)$, is known, the number of
stars in wide binary systems can be determined by integrating $f(a)$ from some
lower cutoff, $a_1$, to the maximum semi-major axis, $a_2$. The power-law
frequency distribution, $f(a)\propto a^\beta$, $\beta=-1$, is known in the
context of wide binaries as the \"Opik distribution (OD; \citealt{oep24}). When
semi-major axis distribution of wide binaries follows the OD, the frequency
distribution of $\log(a)$ is a straight line with a slope of zero \citep{pah07}.
Alternatively, an equivalent representation of OD is the cumulative distribution
$N[<\log(a)]\propto \log(a)$. In this form, OD is a straight line with a
positive slope. We use the cumulative representation, instead of differential,
because it reduces the counting noise in sparsely populated bins (though the
errors become correlated between bins). 

We utilize geometrically-selected candidate binaries (see
Section~\ref{limitations} for a discussion of the kinematic sample), but do not
limit the selection to $\theta < 4\arcsec$, as we did in
Section~\ref{geo_select}. Since $a \propto \theta$, the removal of upper limit
on $\theta$ allows us to probe an extended range of semi-major axes. The
downside is that random pairs dominate at large $\theta$ and a careful
accounting for contamination as a function of $\theta$ is required before the
the $f(a)$ distribution can be constrained. Since we only know the projected
separation of our pairs, we use a statistical relation to calculate the average
semi-major axis, $\langle a\rangle$, as $\langle a\rangle=1.411\, \theta\, d$,
where $d$ is the heliocentric distance \citep{cou60}. Hereafter, we drop the
brackets and simply note the average semi-major axis as $a$.

Figure~\ref{loga_cum} ({\em top left}) shows the cumulative distribution of
$\log(a)$ for candidate wide binaries with $|\delta|<0.2$ selected from the
$0.7< Z/kpc < 1.0$ range. The cumulative distribution does not follow a straight
line, as predicted by the OD, but actually increases its slope with $\log(a)$.
We assume that this is due to an increasing fraction of random pairs at high
$\log(a)$, and proceed to verify this assumption.

Figure~\ref{loga_cum} ({\em top right}) shows the differential distribution of
angular separation for the selected sample. For $\theta>\theta_{max}$ the
observed and random distributions closely match, demonstrating that random pairs
dominate at high $\theta$ (or high $\log(a)$). To calculate how the fraction of
true binaries (or random pairs) changes as a function of $\theta$, we fit
$f_{rnd}(\theta)= C \, \theta$ to the observed histogram, and calculate the
fraction or true binaries, $\epsilon$, using Equation~\ref{eff2} (see
Section~\ref{geo_select}). The calculated $\epsilon$ values, as well as the
best-fit second-degree polynomial, $\epsilon(\theta)$, are shown in
Figure~\ref{loga_cum} ({\em middle left}). As an independent test, the selection
efficiency was calculated using Equation~\ref{eff} (i.e., from the $\delta$
distribution) for three $\theta$-selected subsamples, and the obtained values
agree with $\epsilon(\theta)$ at a level of 1\%. The angular separation for
which $\epsilon$ falls below $\sim5\%$ is defined as $\theta_{max}$. The
fraction of true binaries ($\epsilon(\theta)$) also changes as a function of
$Z$, and is determined separately for different distance bins.

Since the candidates are restricted in $Z$ ($Z_{min}=0.7$ kpc to $Z_{max}=1.0$
kpc in this example) and $\theta$ ($3\arcsec$ to $\theta_{max}$), to ensure a
uniform selection in the $Z$ vs.~$a$ space we define
\begin{equation}
a_{min} = 3\arcsec \cdot 1.411 \cdot 1000 \, Z_{max} \label{a_min}
\end{equation}
and
\begin{equation}
a_{max} = \theta_{max} \cdot 1.411 \cdot 1000 \, Z_{min}/\sin(45\arcdeg)\label{a_max}
\end{equation}
as the minimum and maximum probed semi-major axis, shown as the selection box in
Figure~\ref{loga_cum} ({\em middle right}). The $\sin(45\arcdeg)$ factor is to
account for the fact that the candidates are restricted to high ($b>45\arcdeg$)
Galactic latitudes.

To correct the cumulative distribution of $\log(a)$, we assign a probability
$\epsilon(\theta)$ to each candidate binary in the $a_{min}$ to $a_{max}$ range,
and add the probabilities (instead of counting candidates) when making the
cumulative $\log(a)$ distribution. The corrected cumulative distribution, shown
in Figure~\ref{loga_cum} ({\em bottom left}), follows a straight line up to the
turnover point, $a_{break}$. We define $a_{break}$ as the average semi-major
axis for which the straight line fit to the cumulative distribution deviates by
more than $1.5\%$. In addition to $a_{break}$, we also measure the slope of the
cumulative distribution where it follows the straight line. It can be shown that
the slope of the cumulative distribution is equal to the constant of
proportionality, $N_0$, in \"Opik distribution, $f(a) = N_0/a$. The number of
binaries can be calculated by integrating $f(a)$ from $a_1$ to $a_2$, and we
obtain $N_{bin}=N_0 \, \log(a_2/a_1)$. For integration limits we choose
$a_2=a_{break}$ where we assume that systems with semi-major axes greater than
$a_{break}$ are no longer bound, and $a_1 = 100$ AU (since $a_2 \gg a_1$, the
results are not very sensitive to the choice of $a_1$).

The uncertainty in $a_{break}$, shape of $f(a)$ (or power-law index $\beta$),
and number of binaries ($N_{bin}$) are estimated using Monte Carlo simulations.
We find that the uncertainty in $a_{break}$ is less than 0.1 dex, and the error
on the power-law index ($\beta$) is $\la0.1$. The uncertainty in measuring
$N_{bin}$ is about $10\%$. The corrected cumulative $\log(a)$ distribution
obtained from one of these simulations is shown in Figure~\ref{loga_cum}
({\em bottom right}). The semi-major axis distribution of ``true'' binaries in
the simulation sample is $f(a)\propto a^{-0.8}$, and is valid between 100 AU and
$a_{break}=10,000$ AU. The turnover in the distribution happens because there
are no ``true'' binaries above 10,000 AU, only random pairs, similar to what we
observe in real data. This similarity is a strong warning not to over-interpret
the slope of $f(a)$ beyond $a_{break}$.

To estimate the dependence of $\beta$ (shape of $f(a)$), $a_{break}$, and $N_0$
on color, we divide the $0.7 < Z/kpc < 1.0$ sample into three color subsamples
using $(g-i)_1=1.8$ and $(g-i)_2=1.5$ lines. We find that $f(a)$ follows OD in
all three subsamples ($\beta = -1$), and that the average $a_{break}$ is 3.99,
with a 0.07 root-mean-square scatter. The $a_{break}$ for the full
$0.7<Z/kpc<1.0$ sample is 4.02. These results suggest that $a_{break}$ and the
shape of $f(a)$ are {\em independent of color of binaries}. The $N_0$ value, and
subsequently the number of binaries, will depend on the sample's color range.
For the full $0.7<Z/kpc<1.0$ sample, the number of binaries is
\begin{equation}
N_{bin} = (N_0^1 + N_0^2 + N_0^3) \log_{10}(a_2/a_1),
\end{equation}
where $N_0^i$, $i=1,2,3$, are $N_0$ values measured for each color subsample.
Therefore, the number of binaries calculated for a distance bin will change as
the color range changes. Assuming that the $g-i$ color distribution of binaries
does not change with $Z$, we can use the $g-i$ color distribution for the
$0.7<Z/kpc<1.0$ sample (solid line in Figure~\ref{prob_gi} ({\em top})), to
correct for color incompleteness. We also assume that the fraction of binaries
outside the $0.4 < g-i < 2.8$ color range is small. The correct number of
binaries is then
\begin{equation}
N_{bin} = N_0/A[(g-i)_{min},(g-i)_{max}] \,\log_{10}(a_2/a_1),
\end{equation}
where $A[(g-i)_{min},(g-i)_{max}]$ is the area underneath the solid line 
histogram in Figure~\ref{prob_gi} ({\em top}), between $(g-i)_{min}$ and
$(g-i)_{max}$ ($g-i$ color range for a given distance bin).

The estimated systematic error in $a_{break}$ due to the choice of the
$|\delta|$ cut is measured using $|\delta|<0.1$ and $|\delta|<0.4$ samples. We
find that $a_{break}$ changes by $\la0.03$ dex between these samples. This
result suggests that $a_{break}$ is not sensitive to the choice of the
$|\delta|$ cut. Similarly, the change in $a_{break}$ is less than 0.03 dex if
the estimate of $\epsilon(\theta)$ is off by $\pm0.1$ ($\sim10\%$ change) from
the best-fit $\epsilon(\theta)$.

To establish whether semi-major axis distribution follows the OD in other $Z$
bins, we repeat the $f(a)$ and $a_{break}$ measuring procedure on 8 $Z$ bins,
and show the corrected cumulative distributions with best-fit straight lines in
Figure~\ref{loga_cum_fits}. In general, the corrected cumulative distributions
follow a straight line, and then start to deviate from it at $a_{break}$. In the
$0.1<Z/kpc<0.4$ bin we do not see a turnover due to a narrow range of probed
projected separations ($\theta_{max}=16\arcsec$ limits the range to 3193 AU, see
Figure~\ref{theta_eff_0.1_0.4_geo}), and only determine the upper limit on
$a_{break}$.

As the average height above the Galactic plane increases, the $a_{break}$ moves 
to higher values. We investigate this correlation in more detail in
Figure~\ref{logZ_plots} ({\em top left}). The data follow a straight line
$\log(a_{break})=k \, \log(Z[pc])+l$, where $k=0.72\pm0.05$ and
$l=1.93\pm0.15$, or approximately, $a_{break}[AU] = 12,300 \, Z[kpc]^{0.7}$ in
the $0.3 < Z/kpc < 3.0$ range.

It is possible that $a_{break}$ also depends on the cylindrical radius, $R$,
with the Sun at $R_\odot$=8 kpc, and perhaps on the local density of stars,
$\rho$. Because the sample is dominated by stars at high Galactic latitudes, it
is hard to disentangle the $Z$ dependence from the other two effects (the $R$
range is small, and $\rho$ varies strongly with $Z$). We attempt to do so using
the volume-complete $0.7<Z/kpc<1.0$ sample. First we divide this sample into
three subsamples with median Galactic latitudes, $\langle b \rangle$, of
$35\arcdeg$, $49\arcdeg$, and $80\arcdeg$ and determine $a_{break}$ for each
subsample. The best-fit $a_{break}$ varies by $\sim$0.3 dex between the
low-latitude and high-latitude subsample, despite the same median $Z$. When the
$0.7<Z/kpc<1.0$ sample is divided into the Galactic anticenter
($90\arcdeg < l < 270\arcdeg$) and the Galactic center ($l > 270\arcdeg$ or
$l<90\arcdeg$) subsamples, the best-fit $a_{break}$ varies by $\sim$0.1 dex.
These variations suggest that the best-fit $Z$ dependence does not fully
capture the behavior of $a_{break}$. Nevertheless, they are smaller ($\la0.3$
dex) than the observed variation of $a_{break}$ ($\sim$1 dex). 

The spatial distribution of wide binaries can now be compared to the number
density of all stars as a function of height above the Galactic plane. In
Figure~\ref{logZ_plots} ({\em bottom left}) we show that wide binaries closely
follow the spatial distribution of stars, with exponential decline in the number
density as a function of $Z$. The fraction of binaries relative to the number of
all stars, shown in the bottom right panel, changes by only a factor of 2 over a
range of 3 kpc, starting from $0.9\%$ at $Z=500$ pc and declining to $0.5\%$ at
$Z=3000$ pc.

\subsection{The Limitations of the Kinematic Sample}
\label{limitations}

It would be informative to repeat the $f(a)$ and $a_{break}$ analysis using
kinematically-selected binaries, but unfortunately, the apparent incompleteness
of SDSS-POSS proper motion data at $\theta<9\arcsec$ prevents us in doing so. As
shown in Figure~\ref{theta_fit_500pc}, the number of common proper motion pairs
drops sharply below $\theta=9\arcsec$, probably due to blending of close sources
in the POSS data. Because of this $\theta$ cut-off, for the same range in $Z$,
the effective $a_{min}$ for the kinematic sample is three times that of the
geometric sample (where the lower limit on $\theta$ is $3\arcsec$). In the case
of $0.1<Z/kpc<0.4$ sample observed here, the smallest probed semi-major axis
($a_{min}$) is at 5079 AU, well above the $a_{break}$ value of 4534 AU predicted
by the $a_{break} \propto Z^{0.7}$ relation. Since we are outside the range
where OD is valid, we cannot measure where the turnover in $f(a)$ happens, and
cannot determine $a_{break}$ or $N_{bin}$. In all the other $Z$ bins, $a_{min}$
is also above the predicted $a_{break}$ value, and therefore outside the \"Opik
regime.

\section{Discussion and Conclusions}
\label{discussion}

We have presented a novel approach to photometric parallax estimation based on
samples of candidate wide binaries selected from the Sloan Digital Sky Survey
(SDSS) imaging catalog. Our approach uses the fact that binary system's
components are at equal distances and estimates the photometric parallax
relation for main-sequence stars by minimizing the difference of their distance 
moduli. While this method is similar to constraints on photometric parallax
relation obtained from globular clusters in that it does not require absolute
distance estimates, it has the advantage that it extends to redder colors than
available for globular clusters observed by the SDSS, and it implicitly accounts
for the metallicity effects.

The derived best-fit photometric parallax relations represent
metallicity-averaged relations and thus provide an independent confirmation of
relations proposed by J08 in their study of the Galactic structure. An important
result of this work is our estimate of the expected error distribution for
absolute magnitudes determined from photometric parallax relations (a
root-mean-square scatter of $\sim$0.3 mag, see Section~\ref{scatter}), which is
in good agreement with modeling assumptions adopted by J08. The mildly
non-Gaussian error distribution is consistent with both the impact of unresolved
binary stars, and the variation of photometric parallax relation with
metallicity; we are unable to disentangle these two effects. 

The best-fit photometric parallax relations enabled the selection of
high-efficiency samples of disk wide binaries with $\sim22,000$ candidates, that
include about 14,000 true binary systems (efficiency of $\sim2/3$). Using the
photometric measurements and angular distance of the two components, samples
with efficiency exceeding 80\% can be constructed (see Section~\ref{samples}).
Such samples could be used as a starting point to further increase the selection
efficiency with the aid of radial velocity measurements. Spectral observations
of systems where the brighter component is an F/G star, for which it is easy to
estimate metallicity, could be used to calibrate both spectroscopic and
photometric methods for estimating metallicity of cooler K and M dwarfs.
Compared to the state-of-the-art catalogs of wide binaries by \citet{cg04} and 
\citet{lb07}, the samples discussed here represent a significant increase in the
number of potential binaries, and probe larger distances (to $\sim4$ kpc). To
facilitate further studies of wide binaries, we make the catalog publicly
available\footnote{The catalog can be downloaded from
\url[HREF]{http://www.astro.washington.edu/bsesar/SDSS\_wide\_binaries.tar.gz}}.

Using the high-efficiency subsamples, we analyzed their dynamical and physical 
properties. We find that the spatial distribution of wide binaries follows the
distribution of single stars to within a factor of 2, and that the probability
for a star to be in a wide binary system is independent of its color. However, 
given this color, the companions of red components seem to be drawn randomly
from the stellar luminosity function, while blue components have a larger
blue-to-red companion ratio than expected from luminosity function (see
Section~\ref{color}). These results are consistent with recent results by
\citet{lb07}, and provide strong constraints for the scenarios describing the
formation of such systems (e.g., \citealt{gie06} and references therein;
\citealt{cla07}; \citealt{hur07}).

We also study the semi-major axis distribution of wide binaries in the
$2,000-47,000$ AU range (see Section~\ref{spatial}). The observed distribution
is well described by the \"Opik distribution, $f(a)\propto 1/a$, for
$a<a_{break}$, where $a_{break}$ increases roughly linearly with the height
above the Galactic plane ($a_{break}\sim 12,300$ AU at $Z=1$ kpc).
Alternatively, the $a_{break}$ correlates with the local number density of stars
as $a_{break} \propto \rho^{-1/4}$, but we are unable to robustly identify the
dominant correlation ($Z$ and $\rho$ are highly correlated). 

The distribution of semi-major axes for wide binaries was also discussed by
\citet{cg04}. They used a sample of wide binaries selected using common proper 
motion from the rNLTT catalog \citep{gs03}, and found $f(a)\propto 1/a^{1.6}$,
with no evidence of a turnover at $a \la 3000$. Their sample extended to larger
angular separations than ours, and probed smaller distances. On the other hand, 
\citet{pah07} used wide binaries from the same \citet{cg04} sample, and detected
\"Opik distribution, $f(a)\propto 1/a$, for $a<3,000$, consistent with the
result of \citet{cg04}. In a recent study,  L\'epine \& Bongiorno searched for 
faint common proper motion companions of Hipparcos stars and detected a turnover
from \"Opik distribution to a steeper distribution around $a\sim3,000$ AU. Their
sample also probed much smaller distances than ours. We compare these results in
Figure~\ref{Fig:compare}. As evident, the variation of $a_{break}$ with distance
from the Galactic plane detected here (approximately with distance, as shown in
Figure~\ref{Fig:compare}, since stars in our sample are mostly at high galactic
latitudes), is consistent with the above results that are based on more local
samples. In particular, this comparison of different studies suggests that the
flattening of $f(a)$ for small $a$ that ``puzzled'' Chanam\'e \& Gould (see
their section 4.3) is probably due to a combination of selection effects and the
approach of the domain where \"Opik distribution is valid in their sample. 

The \"Opik distribution suggests that the process of star formation produces
multiple stars, which evolve towards binaries after ejecting one or more single
stars \citep{pah07}. The departure from the \"Opik distribution may be evidence
for disruption of wide binaries over long periods of time by passing stars,
giant molecular clouds, massive compact halo objects (MACHOs), or disk and
Galactic tides \citep{heg75,wsw87,ycg04}. However, we note that the
$a_{break} \propto \rho^{-1/4}$ correlation (see Figure~\ref{logZ_plots}) is
outside the expected range discussed by \citet{ycg04}
($a_{break}\propto \rho^{-2/3}$ for close strong encounters, and
$a_{break}\propto \rho^{-1}$ for weak encounters). 

The samples presented here can be further refined and enlarged. First, the SDSS
covers only a quarter of the sky. Upcoming next-generation surveys, such as the
SkyMapper \citep{kel07}, the Dark Energy Survey \citep{fla07}, Pan-STARRS
\citep{kai02} and the Large Synoptic Survey Telescope (\citealt{ive08b}, LSST
hereafter), will enable the construction of such samples over most of the sky.
Due to fainter flux limits (especially for the Pan-STARRS and LSST), the samples
will probe a larger distance range and will reach the halo-dominated parts of
the Galaxy. Furthermore, due to improved photometry and seeing (e.g., for the
LSST, by about a factor of two), the selection will be more robust. We scale the
20,000 candidates discussed here, assuming $log(N) = C + 0.4\,r$, to the LSST
depth that enables accurate photometric metallicity ($r<23$; I08a) and predict a
minimum sample size of $\sim$400,000 candidate wide binary systems in 20,000
deg$^2$ of sky. It is likely that the sample would include more than a million
systems due to the increase of the stellar counts close to the Galactic plane.

Another important development will come from the Gaia mission
\citep{per01,wil05}, which will provide direct trigonometric distances for stars
with $r<20$. With trigonometric distances, accurate photometric parallax
relation can be used to provide strong constraints on the incidence and color
distribution of unresolved multiple systems. Until then, a radial velocity 
survey of candidate binaries assembled here could help with pruning the sample 
from random associations, and with better characterization of various
selection effects. 

\acknowledgments

This work was supported by the NSF grant AST-0707901, and the NSF grant
AST-0551161 to the LSST for design and development activity. M.~J.~gratefully
acknowledges support from the Taplin Fellowship and from the NSF grant
PHY-0503584. We are grateful to Nick Cowan and Eric Agol (UW) for help with the
Markov chain Monte Carlo code. Early motivation for this analysis came from a
preliminary study by Taka Sumi.

Funding for the SDSS and SDSS-II has been provided by the Alfred P. Sloan
Foundation, the Participating Institutions, the National Science Foundation, the
U.S. Department of Energy, the National Aeronautics and Space Administration,
the Japanese Monbukagakusho, the Max Planck Society, and the Higher Education
Funding Council for England. The SDSS Web Site is http://www.sdss.org/.

The SDSS is managed by the Astrophysical Research Consortium for the
Participating Institutions. The Participating Institutions are the American
Museum of Natural History, Astrophysical Institute Potsdam, University of Basel,
University of Cambridge, Case Western Reserve University, University of Chicago,
Drexel University, Fermilab, the Institute for Advanced Study, the Japan
Participation Group, Johns Hopkins University, the Joint Institute for Nuclear
Astrophysics, the Kavli Institute for Particle Astrophysics and Cosmology, the
Korean Scientist Group, the Chinese Academy of Sciences (LAMOST), Los Alamos
National Laboratory, the Max-Planck-Institute for Astronomy (MPIA), the
Max-Planck-Institute for Astrophysics (MPA), New Mexico State University, Ohio
State University, University of Pittsburgh, University of Portsmouth, Princeton
University, the United States Naval Observatory, and the University of
Washington.

\appendix

\section{SQL Queries}
\label{appendix}

The following SQL queries were used to select initial samples of candidate
binaries through the SDSS CasJobs interface. When running these queries, the
database context must be set to ``DR6'' or higher.

\begin{verbatim}
select -- geometric selection of candidate binaries

round(p1.ra,6) as ra1, round(p1.dec,6) as dec1, round(p1.extinction_r,3) as rExt1,
round(p1.psfMag_u,3) as psf_u1, round(p1.psfMag_g,3) as psf_g1,
round(p1.psfMag_r,3) as psf_r1, round(p1.psfMag_i,3) as psf_i1,
round(p1.psfMag_z,3) as psf_z1, round(p1.psfMagErr_u,3) as psfErr_u1,
round(p1.psfMagErr_g,3) as psfErr_g1, round(p1.psfMagErr_r,3) as psfErr_r1,
round(p1.psfMagErr_i,3) as psfErr_i1, round(p1.psfMagErr_z,3) as psfErr_z1,
p1.objid as objid1,

round(p2.ra,6) as ra2, round(p2.dec,6) as dec2, round(p2.extinction_r,3) as rExt2,
round(p2.psfMag_u,3) as psf_u2, round(p2.psfMag_g,3) as psf_g2,
round(p2.psfMag_r,3) as psf_r2, round(p2.psfMag_i,3) as psf_i2,
round(p2.psfMag_z,3) as psf_z2, round(p2.psfMagErr_u,3) as psfErr_u2,
round(p2.psfMagErr_g,3) as psfErr_g2, round(p2.psfMagErr_r,3) as psfErr_r2,
round(p2.psfMagErr_i,3) as psfErr_i2, round(p2.psfMagErr_z,3) as psfErr_z2,
p2.objid as objid2,

round(NN.distance*60,3) as theta

into mydb.binaryClose

from Neighbors as NN join star as p1 on p1.objid = NN.objid
join star as p2 on p2.objid = NN.neighborobjid
where NN.mode = 1 and NN.neighbormode = 1
and NN.type = 6 and NN.neighbortype = 6

and p1.psfMag_r between 14 and 20.5
and (p1.flags_g & '229802225959076') = 0 and (p1.flags_r & '229802225959076') = 0
and (p1.flags_i & '229802225959076') = 0 and (p1.flags_g & '268435456') > 0
and (p1.flags_r & '268435456') > 0 and (p1.flags_i & '268435456') > 0

and p2.psfMag_r between 14 and 20.5
and (p2.flags_g & '229802225959076') = 0 and (p2.flags_r & '229802225959076') = 0
and (p2.flags_i & '229802225959076') = 0 and (p2.flags_g & '268435456') > 0
and (p2.flags_r & '268435456') > 0 and (p2.flags_i & '268435456') > 0

and (p1.psfMag_r-p1.extinction_r) < (p2.psfMag_r-p2.extinction_r)
and (p1.psfMag_g-p1.extinction_g - p1.psfMag_i+p1.extinction_i)<
(p2.psfMag_g-p2.extinction_g - p2.psfMag_i+p2.extinction_i)

and NN.distance*60 between 3 and 4
\end{verbatim}

\begin{verbatim}
select -- kinematic selection of candidate binaries

round(p1.ra,6) as ra1, round(p1.dec,6) as dec1, round(p1.extinction_r,3) as ext1,
round(p1.psfMag_u,3) as u1, round(p1.psfMag_g,3) as g1,
round(p1.psfMag_r,3) as r1, round(p1.psfMag_i,3) as i1,
round(p1.psfMag_z,3) as z1, round(p1.psfMagErr_u,3) as uErr1,
round(p1.psfMagErr_g,3) as gErr1, round(p1.psfMagErr_r,3) as rErr1,
round(p1.psfMagErr_i,3) as iErr1, round(p1.psfMagErr_z,3) as zErr1,
(case when ((p1.flags & '16') = 0) then 1 else 0 end) as ISOLATED1,
NN.objid as objid1,

round(p2.ra,6) as ra2, round(p2.dec,6) as dec2, round(p2.extinction_r,3) as ext2,
round(p2.psfMag_u,3) as u2, round(p2.psfMag_g,3) as g2,
round(p2.psfMag_r,3) as r2, round(p2.psfMag_i,3) as i2,
round(p2.psfMag_z,3) as z2, round(p2.psfMagErr_u,3) as uErr2,
round(p2.psfMagErr_g,3) as gErr2, round(p2.psfMagErr_r,3) as rErr2,
round(p2.psfMagErr_i,3) as iErr2, round(p2.psfMagErr_z,3) as zErr2,
(case when ((p2.flags & '16') = 0) then 1 else 0 end) as ISOLATED2,
NN.neighborobjid as objid2,

round(NN.distance*60,3) as theta,
round(s1.pmL,3) as pmL1, round(s1.pmB,3) as pmB1,
round(s2.pmL,3) as pmL2, round(s2.pmB,3) as pmB2

into mydb.binaryPM

from Neighbors as NN join star as p1 on p1.objid = NN.objid
join star as p2 on p2.objid = NN.neighborobjid
join propermotions as s1 on s1.objid = NN.objid
join propermotions as s2 on s2.objid = NN.neighborobjid

where NN.mode = 1 and NN.neighbormode = 1
and NN.type = 6 and NN.neighbortype = 6

and p1.psfMag_r between 14 and 19.5
and (p1.flags_g & '229802225959076') = 0 and (p1.flags_r & '229802225959076') = 0
and (p1.flags_i & '229802225959076') = 0 and (p1.flags_g & '268435456') > 0
and (p1.flags_r & '268435456') > 0 and (p1.flags_i & '268435456') > 0

and p2.psfMag_r between 14 and 19.5
and (p2.flags_g & '229802225959076') = 0 and (p2.flags_r & '229802225959076') = 0
and (p2.flags_i & '229802225959076') = 0 and (p2.flags_g & '268435456') > 0
and (p2.flags_r & '268435456') > 0 and (p2.flags_i & '268435456') > 0

and (p1.psfMag_r-p1.extinction_r) < (p2.psfMag_r-p2.extinction_r)
and (p1.psfMag_g-p1.extinction_g - p1.psfMag_i+p1.extinction_i)<
(p2.psfMag_g-p2.extinction_g - p2.psfMag_i+p2.extinction_i)

and s1.match = 1 and s2.match = 1
and s1.sigra < 350 and s1.sigdec < 350
and s2.sigra < 350 and s2.sigdec < 350
and sqrt(power(s1.pmL - s2.pmL,2) + power(s1.pmB - s2.pmB,2)) < 5
and (case when sqrt(power(s1.pmL,2) + power(s1.pmB,2)) >
sqrt(power(s2.pmL,2) + power(s2.pmB,2)) then
sqrt(power(s1.pmL,2) + power(s1.pmB,2)) else
sqrt(power(s2.pmL,2) + power(s2.pmB,2)) end) between 15 and 400
\end{verbatim}

\section{  The Limitations of the Reduced Proper Motion Diagram   }

Recent analysis of metallicity and kinematics for halo and disk stars by I08a
provides sufficient information to understand the behavior of the reduced proper
motion diagram in quantitative detail (including both the sequence separation
and their widths), and to demonstrate that its efficiency for separating halo
and disk stars deteriorates at distances beyond a few kpc from the Galactic
plane. As Equation~\ref{Eq:rpm2} shows, for a population of stars with the same
$v_t$, the reduced proper motion is a measure of their absolute magnitude. For
two stars with the same color that is sensitive to the effective temperature
(such as the $g-i$ color), but with different metallicities and tangential
velocities, the difference in their reduced proper motions is
\begin{equation}
\label{Eq:rpm3}
 \Delta r_{RPM} = r^H_{RPM} - r^D_{RPM} =  \Delta M_r + 5\log{\left({v^H_t \over v^D_t}\right)},
\end{equation}
where $H$ and $D$ denote the two stars. In the limit that the {\it shape} of the
photometric parallax relation does not depend on metallicity, $\Delta M_r$ does
not depend on color, and is fully determined by the metallicity difference of
the two stars (or populations of stars). Using metallicity distributions for
disk and halo stars obtained by I08a, and their expression for
$\Delta M_r([Fe/H])$ (Equation~A2), we find that the expected offset between
$M_r$ for halo and disk stars with the same $g-i$ color varies from 0.6 mag for
stars at 1 kpc from the Galactic plane to 0.7 mag at 5 kpc from the plane, where
the variation is due to the vertical metallicity gradient for disk stars. The
finite width of halo and disk metallicity distributions induces a spread of
$M_r$ (root-mean-square scatter computed using interquartile range) of 0.15 mag
for disk stars and 0.18 mag for halo stars. 

The effect of metallicity on the separation of halo and disk sequences in the
reduced proper motion diagram is smaller than the effect of different tangential
velocity distributions. Assuming for simplicity that stars are observed towards
a Galactic pole, and that the median heliocentric tangential velocities are 30
km s$^{-1}$ for disk stars and 200 km s$^{-1}$ for halo stars, the induced
separation of their reduced proper motion sequences is $\sim$4.1 mag (the
expected scatter in the reduced proper motion due to finite velocity dispersion
is $\sim$1-1.5 mag). Together with the $\sim0.7$ mag offset due to different
metallicity distributions, the separation of $\sim$5 mag between the two
sequences makes the reduced proper motion diagram a promising tool for
separating disk and halo stars.

However, the reduced proper motion diagram is an efficient tool only for stars
within 1-2 kpc from the Galactic plane. The main reason for this limitation is
the decrease of rotational velocity for disk stars with distance from the
Galactic plane, with a gradient of about $-30$ km s$^{-1}$ kpc$^{-1}$ (see
Section 3.4.2 in I08a). As the difference in rotational velocity between halo
and disk stars diminishes with the distance from the plane, the separation of
their reduced proper motion sequences decreases, too. A mild increase in the
velocity dispersion of disk stars, as well as a decrease of their median
metallicity with the distance from the plane, also decrease the sequence
separation, but the dominant cause is the rotational velocity gradient.

To illustrate this effect, we select a sample of $\sim$60,000 stars with
$14<r<20$ and $0.2<g-r<0.4$, that are observed towards the north Galactic pole
($b>70\arcdeg$). In this color range it is possible to separate disk and halo
stars using photometric metallicity estimator from I08a, and we further select a
sample of $\sim$16,000 likely disk stars with $[Fe/H] > -0.9$, and a sample of
$\sim$34,400 likely halo stars with $[Fe/H] < -1.1$ (see Figure~9 in I08a for
justification). Their proper motion distributions as functions of distance from
the Galactic plane, $Z$, are shown in the top left panel in
Figure~\ref{Fig:App1}. Because of the gradient in the rotational velocity for
disk stars, their median proper motion becomes constant at $\sim$8 mas yr$^{-1}$
beyond $Z\sim2$ kpc, while the median proper motion for halo stars is roughly
proportional to $1/Z$, with a value of $\sim$11 mas yr$^{-1}$ at $Z=5$ kpc.

The top right panel in Figure~\ref{Fig:App1} shows the positions and widths of
the reduced proper motion sequences for disk and halo stars as functions of $Z$,
and the two bottom panels show the sequence cross-sections for stars with
$Z=1-1.5$ kpc and $Z=3.5-4$ kpc. At distances beyond $\sim$2 kpc from the plane,
the reduced proper motion diagram ceases to be an efficient tool for separating
halo and disk stars because the two sequences start to significantly overlap.
This increasing overlap is a result of the rotational velocity gradient for disk
stars, and the finite width of halo and disk velocity distributions, and would
be present even for {\it infinitely accurate} measurements (with the proper
motion errors of $\sim$3 mas yr$^{-1}$ per coordinate, \citealt{mun04}, the
sequence widths of $\sim$1.0-1.5 mag are dominated by velocity dispersions).
Hence, beyond $\sim$2 kpc from the plane, metallicity measurements are necessary
to reliably separate disk and halo populations.

The above analysis is strictly valid only for fields towards the north Galactic
pole. \citet{sg03} found that the position of disk and halo reduced proper
motion sequences, relative to their positions at the north Galactic pole, varies
with galactic latitude as 
\begin{equation}
\label{Eq:rpm4}
 \Delta r_{RPM}(b) = 5\log(v_t/v_t^{NGP}) = -1.43 \left(1-\sin(|b|)\right),
\end{equation}
where $v_t^{NGP}$ is the median value of $v_t$ for stars observed towards the
north Galactic pole. This result is a bit unexpected because it does not contain
longitudinal variation due to projection effects of the rotational motion of the 
local standard of rest. We show the variation of $\Delta r_{RPM}$, for stars
with $0.2<g-r<0.4$, as a function of galactic coordinates in
Figure~\ref{Fig:App2}. We use photometric metallicity to separate stars into
disk and halo populations. As figure demonstrates, the longitudinal dependence
is present for halo sample, but not for disk samples. We have generated
simulated behavior of $\Delta r_{RPM}$ using kinematic model from I08a, and
reproduced the observed behavior to within the measurement noise. It turns out
that the vertical gradient of rotational velocity for disk stars is fully
responsible for the observed strong dependence of $\Delta r_{RPM}$ on latitude,
and which masks the dependence on longitude. Hence, the $\sin(|b|)$ term
proposed by \citet{sg03} is an indirect discovery of the vertical gradient of
rotational velocity for disk stars! These empirical models also show that a
linear dependence of $\Delta r_{RPM}(b)$ on $\sin(|b|)$ is only approximately
correct, and that it ignores the dependence on distance. While a more involved
best-fit expression is possible (full two-dimensional consideration of proper
motion also helps to better separate disk and halo stars), we find that halo
stars can always be efficiently rejected at $|b|>30\arcdeg$, if the separator
shown in Figure~\ref{Fig:1wd} is shifted upwards by 1 mag. 

\section{The Modeling of Unresolved Binaries in the Samples of Wide Binaries}
\label{model}

One major uncertainty when using a photometric parallax relation is the lack of
information whether the observed ``star'' is a single star, or a binary
(multiple) system. If the observed ``star'' is a binary system, its luminosity
will be underestimated, with the magnitude of the offset depending on the actual
composition of the binary. To model this offset, or to correct for it, one would
ideally like to have a probability density map that gives the probability of a
magnitude offset, $\Delta M_r$, as a function of the {\em observed} binary
system's color.

To construct such a map, we have generated a sample of 100,000 unresolved binary
systems by randomly pairing stars drawn from the \citet{ktg90} luminosity
function. By independently drawing the luminosities of each component to 
generate unresolved binary systems, we implicitly assume that the formation of
each component is unaffected by the presence of the other. While there are other
proposed mechanisms for binary formation (\citealt{cla07}, and references
therein), we have chosen this one because it was easy to implement.

For every unresolved binary system we calculate the total $r$ band luminosity,
and the $r-i$ and $g-i$ color of the system. The magnitude offset, $\Delta M_r$,
caused by unresolved binarity, is obtained as the difference between the true
$r$ band absolute magnitude, and the absolute magnitude for the pair's joint
$r-i$ color calculated using Equation~\ref{Mr_bright}. The probability density
map is then simply the number of unresolved binary systems (normalized with the
total number of systems at a given color) as a function of $\Delta M_r$ and
pair's joint $g-i$ color, shown in Figure~\ref{dMr_gi}.

It is worth noting that, with the adopted binary formation mechanism, the
magnitude offset is the smallest ($\Delta M_r<0.1$ mag) for the bluest stars,
and greatest ($\Delta M_r>0.7$ mag) for the reddest stars. Because of this, the
scatter due to unresolved binarity in the $\delta$ distribution should be more
pronounced in a sample of red stars ($g-i>2.0$), than in a sample of blue stars.

The map shown in Figure~\ref{dMr_gi} can be parametrized as a Gaussian
distribution $P(\Delta M_r|\mu, \sigma)$, where
\begin{equation}
\mu = 0.037+0.10(g-i)+0.09(g-i)^2-0.012(g-i)^3
\end{equation}
is the median $\Delta M_r$, and
\begin{equation}
\sigma = 0.041+0.03(g-i)+0.15(g-i)^2-0.057(g-i)^3
\end{equation}
is the scatter (determined from the interquartile range). To verify the validity
of this parametrization, we subtract $\Delta M_r$ and $\mu$, normalize the
difference with $\sigma$, find the distribution of such values, and fit a
Gaussian to it. As shown in Figure~\ref{dMr_chi_gauss}, the distribution of
normalized residuals is well described by a Gaussian with $\sigma = 0.9$. The
two peaks in the distribution are due to highly asymmetric distributions of
$\Delta M_r$ values around the median $\Delta M_r$ for the bluest ($g-i\sim0.1$)
and reddest ($g-i\sim2.9$) systems.

To create a sample of wide binaries where some of the stars are unresolved
binary systems, first we select pairs with $20\arcsec<\theta<30\arcsec$ from
the initial sample of stellar pairs. Following the procedure described in
Section~\ref{algorithm}, we create the ``true'' wide binaries by changing the
$r_2$ magnitude using Equation~\ref{r2_mag}, and add 0.15 mag of Gaussian noise
to simulate the scatter in the photometric parallax due to photometric errors.
A fraction of stars is then randomly converted to unresolved binary systems by
subtracting a $\Delta M_r$ value from the $r$ band (apparent) magnitude, where
the $\Delta M_r$ is drawn from a $g-i$ color-dependent
$P(\Delta M_r|\mu, \sigma)$ distribution.

Figure~\ref{gauss_fit_red} shows the $\delta$ distribution for such a mock
sample, where the components are redder than $g-i=2.0$ and have a 40\%
probability to be unresolved binary systems. Different configurations of single
stars and unresolved binaries that contribute to the observed $\delta$
distribution can be easily identified. Wide binaries where both components are
single stars contribute the central narrow Gaussian, with its width due to
photometric errors. If the brighter component is an unresolved binary system,
its absolute magnitude is underestimated, and the result is an offset in
$\delta$ in the negative direction. A similar outcome happens if the fainter
component is an unresolved binary system, but the offset is positive. Single
star-unresolved binary configurations, therefore, contribute the left and the
right Gaussians. If both components are unresolved binary systems, the $\delta$
will be centered on zero and will be $\sigma_0 \sqrt2$ wide, where $\sigma_0$ is
the width of the $(\Delta M_r - \mu)$ distribution. This behavior is consistent
with the $\delta$ distributions observed in Figure~\ref{delta_hists}.

\clearpage


\begin{deluxetable*}{crrrr}
\tabletypesize{\scriptsize}
\tablecolumns{5}
\tablewidth{0pc}
\tablecaption{The centers, widths, and areas for best-fit Gaussian distributions\label{gauss}}
\tablehead{
\multicolumn{1}{c}{} & \multicolumn{2}{c}{Geometrically-selected sample} & \multicolumn{2}{c}{Kinematically-selected sample} \\
\cline{1-5} \\
\colhead{ } & \colhead{Narrow Gaussian} & \colhead{Wide Gaussian} &
\colhead{Narrow Gaussian} & \colhead{Wide Gaussian} 
}
\startdata
Center                            &  -0.01 &   -0.03 &  -0.05 &   0.01 \\
Width                             &   0.12 &    0.54 &   0.11 &   0.51 \\
Area\tablenotemark{a}             &   0.26 &    0.74 &   0.34 &   0.66
\enddata
\tablenotetext{a}{Areas of the narrow and wide Gaussians sum to 1}
\end{deluxetable*}

\clearpage

\begin{deluxetable*}{crrr}
\tabletypesize{\scriptsize}
\tablecolumns{4}
\tablewidth{0pc}
\tablecaption{The conditional probability density functions $P[(g-i)_B|(g-i)_A]=a+b(g-i)+c(g-i)^2$\label{tbl_wide_bin_prob}}
\tablehead{
\multicolumn{1}{c}{} & \multicolumn{3}{c}{Best-fit parameters} \\
\cline{1-4} \\
\colhead{$(g-i)_A$ bin} & \colhead{a} & \colhead{b} & \colhead{c} 
}
\startdata
$0.4 < (g-i)_A < 0.8$ & 0.38 &     0 &    0 \\
$0.8 < (g-i)_A < 1.2$ & 0.46 &     0 &    0 \\
$1.2 < (g-i)_A < 1.6$ & 0.37 &     0 &    0 \\
$1.6 < (g-i)_A < 2.0$ & 0.37 &     0 &    0 \\
$2.0 < (g-i)_A < 2.4$ & 0.08 &  0.14 & 0.04 \\
$2.4 < (g-i)_A < 2.8$ & 0.23 & -0.50 & 0.38
\enddata
\end{deluxetable*}

\clearpage


\clearpage

\clearpage

\clearpage

\clearpage

\begin{figure}
\epsscale{1.0}
\plotone{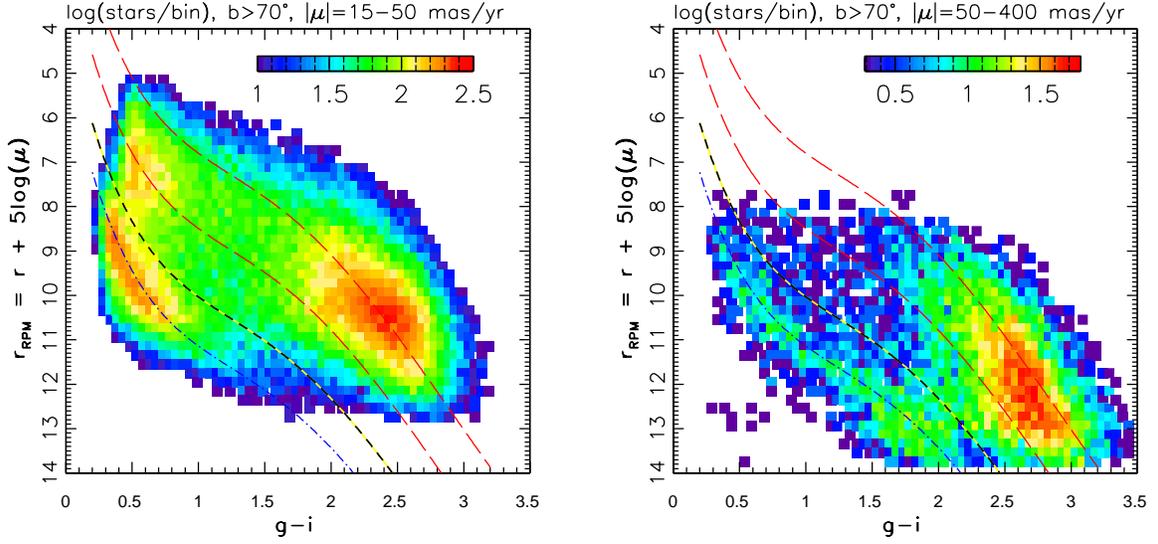}
\caption{The reduced proper motion diagrams for two subsamples of stars shown in
Fig.~\ref{Fig:DMvsColor}. The color-coded maps show the logarithm of the number
of stars per pixel, according to the legends. The left panel corresponds to a
sample of $\sim$446,000 stars with proper motions in the range 15-50 mas/yr, and
the right panel to a sample of 43,000 stars from the range 50-400 mas/yr. The
requirement of larger proper motions introduces bias towards closer, and thus
redder stars. Two two long-dashed lines in each panel correspond to photometric
parallax relation from I08a, evaluated for $[Fe/H]=-0.6$ and with tangential
velocity of 55 km/s (top curve) and 120 km/s (bottom curve). This variation of
tangential velocity is consistent with the rotational velocity gradient
discussed by I08a. The dot-dashed line is evaluated for $[Fe/H]=-1.5$ and with
tangential velocity of 300 km/s. The short-dashed line (second from the bottom)
separates disk and halo stars, and is evaluated for $[Fe/H]=-1.5$ and with
tangential velocity of 180 km/s.
\label{Fig:1wd}}
\end{figure}

\clearpage

\begin{figure}
\epsscale{0.85}
\plotone{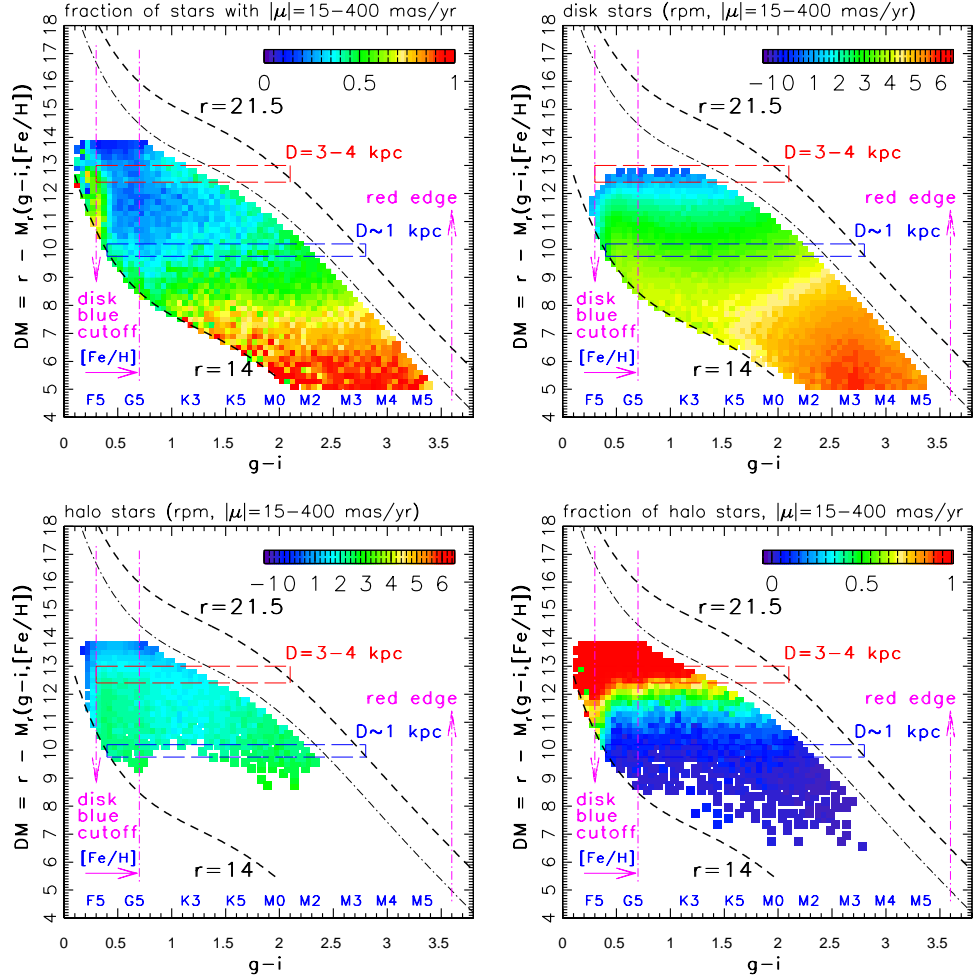}
\caption{Analogous to Fig.~\ref{Fig:DMvsColor}, for subsamples selected using
proper motion measurements. Out of 2.8 million stars shown in
Fig.~\ref{Fig:DMvsColor}, 1.24 million are brighter than $r=19.5$ and have
proper motion measurements. Of those, 498,000 have proper motion in the range
15-400 mas/yr (only 10\% of selected stars have proper motions greater than 50
mas/yr). The color-coded map in the top left panel shows the fraction of such
stars, as a function of distance and the $g-i$ color. At a distance of $\sim$1
kpc, about half of all stars have proper motion larger than 15 mas/yr. The top
right panel shows the counts of candidate disk stars, selected as stars above
the separator shown in Fig.~\ref{Fig:1wd}, and the bottom left panel shows halo
stars selected from below the separator. The bottom right panel shows the counts
of halo stars, as a fraction of all stars selected using the reduced proper
motion diagram. Note that beyond 3 kpc, the sample is dominated by halo stars.
\label{Fig:2wd}}
\end{figure}

\clearpage

\begin{figure}
\epsscale{0.45}
\plotone{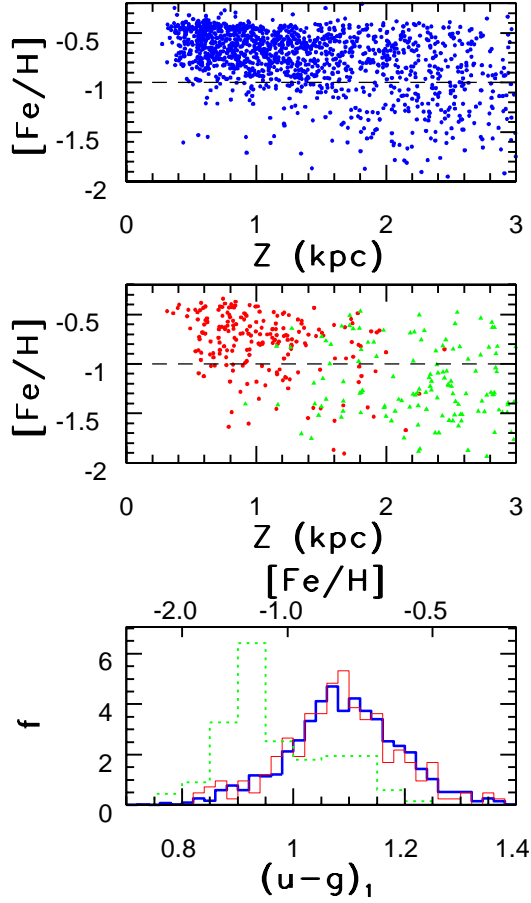}
\caption{
{\em Top:} The photometric metallicity vs.~distance from the Galactic plane
diagram for candidate binaries selected from the geometric sample using
$|\delta|<0.4$ and $0.2<(g-r)_1<0.4$. The $|\delta|<0.4$ cut is used to reduce
the contamination by random pairs (see Section~\ref{scatter}). Note that the
fraction of low-metallicity halo binaries ($[Fe/H]<-1$) becomes significant only
at $Z>2$ kpc. {\em Middle:} Analogous to the top panel, except that binaries
from the kinematic sample are shown. Dots correspond to binaries with reduced
proper motions characteristic of disk binaries, and triangles to candidate halo
binaries. Note that binaries with disk-like metallicity ($[Fe/H]>-1$) at large
distances ($Z>2$ kpc) are misclassified as halo binaries. {\em Bottom:} The
comparison of the $(u-g)_1$ color distributions, and corresponding photometric
metallicity distributions, for binaries from the top two panels. The metallicity
vs.~$u-g$ color transformation is taken from I08a. The distribution for binaries
from the geometric sample is shown by the thick solid line, and the
distributions for binaries from the kinematic sample are shown by the thin solid
line (disk candidates) and dotted line (halo candidates). 
\label{plot_ug}}
\end{figure}

\clearpage

\begin{figure}
\epsscale{0.55}
\plotone{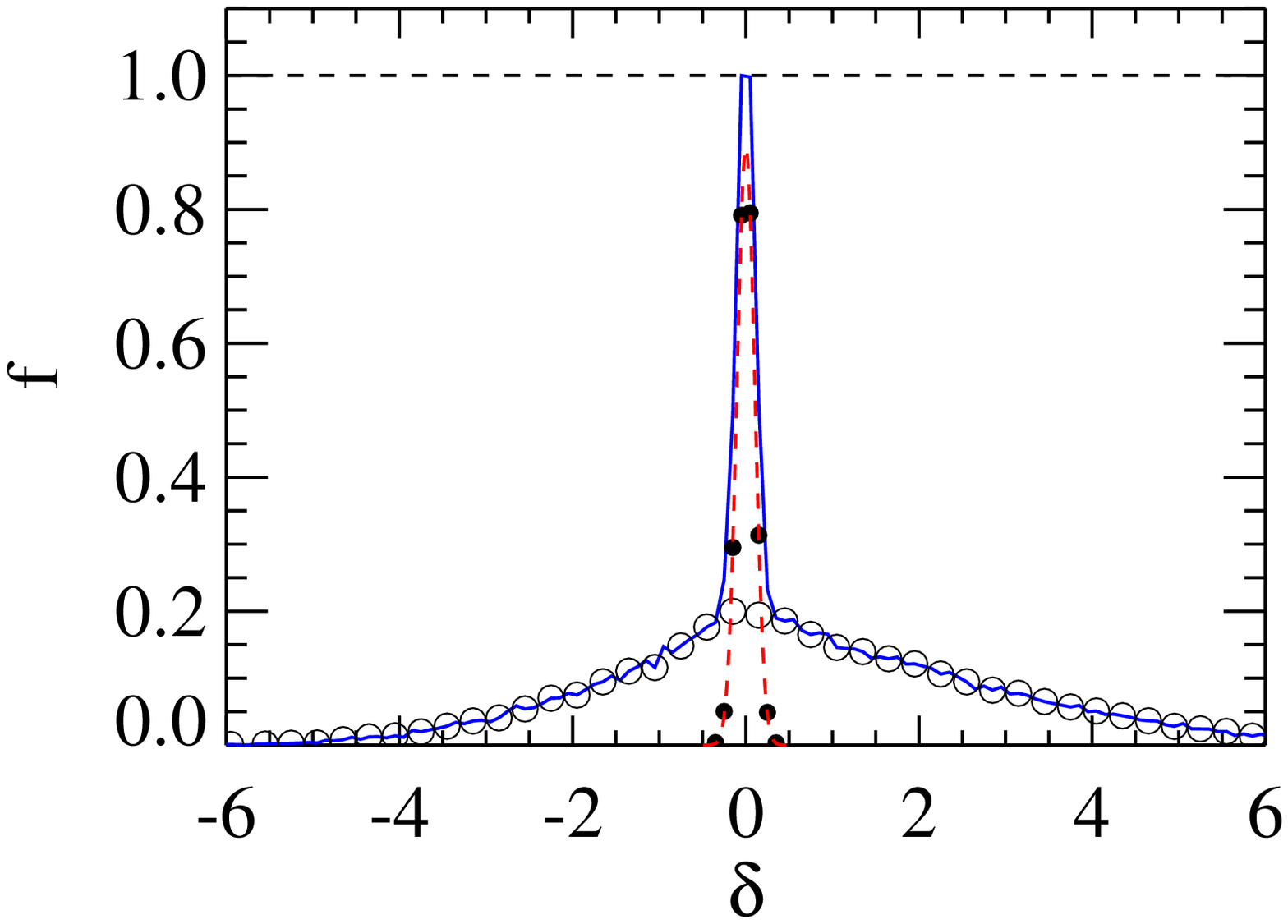}

\plotone{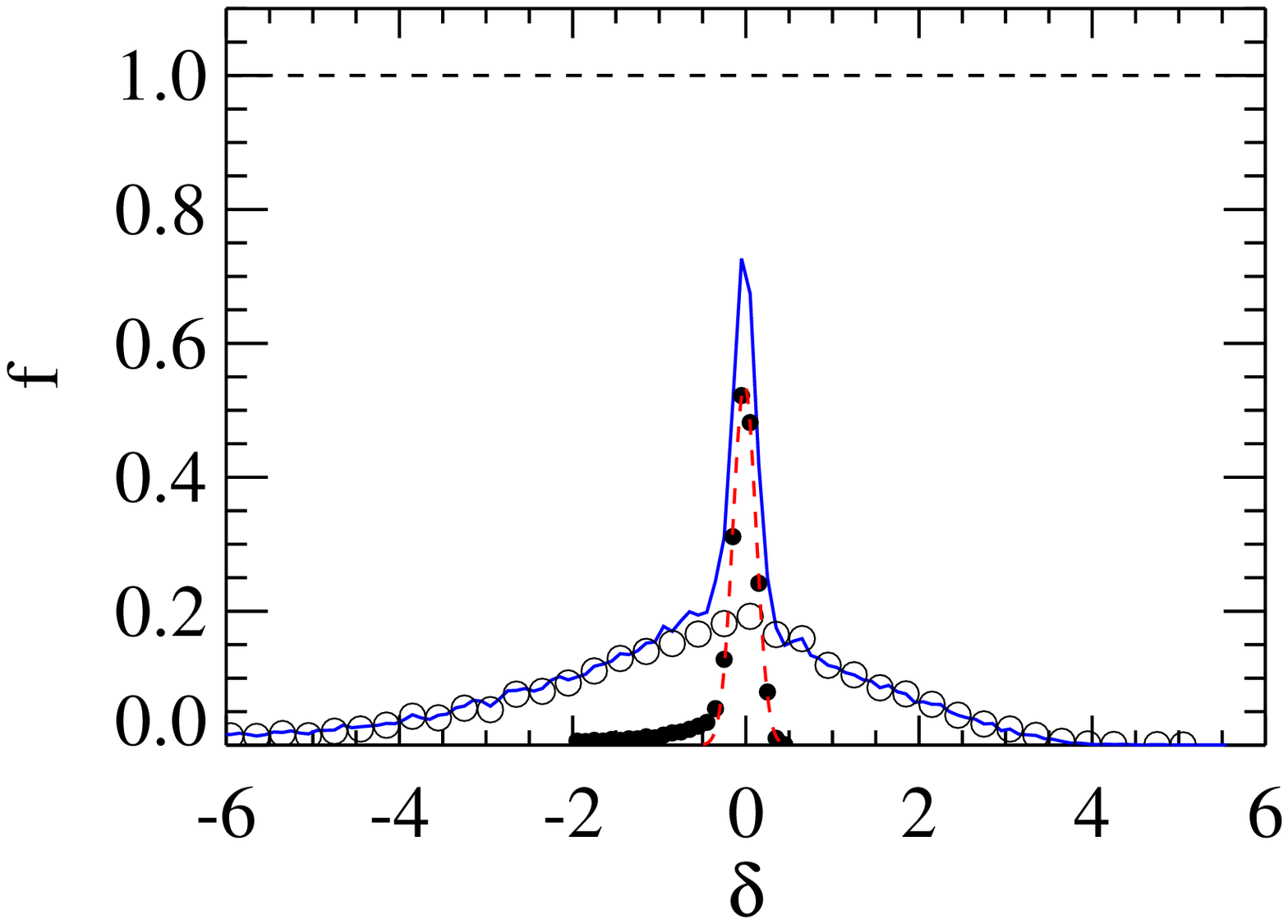}
\caption{
Distribution of $\delta=(M_{r2}-M_{r1})-(r_2-r_1)$ values for a mock sample of
candidate binaries ({\em solid line}) when $M_r=M_r(r-i|\mathbf{p_0})$
({\em top}), and for a $M_r$ different from $M_r(r-i|\mathbf{p_0})$
({\em bottom}). The fraction of random pairs (the contamination) in the sample
is $80\%$. The $\delta$ distribution for ``true'' binaries ({\em dots}) is
obtained by subtracting the $\delta$ distribution of random pairs ({\em open
circles}) from the candidate binary $\delta$ distribution. The best-fit Gaussian
for the ``true'' binaries $\delta$ distribution is centered on 0 and 0.1 mag
wide in the top panel, and centered on -0.02 and 0.13 mag wide in the bottom
panel.
\label{delta_hist_mock}}
\end{figure}

\clearpage

\begin{figure}
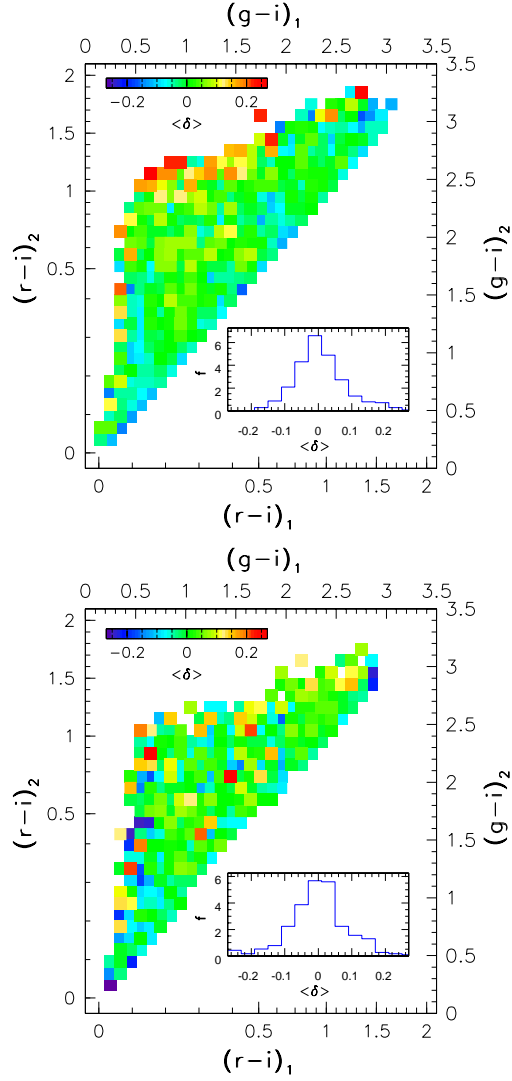

\epsscale{0.44}
\plotone{f9a.eps}

\plotone{f9b.eps}
\caption{
The dependence of median $\delta$, $\langle \delta \rangle$, values on $r-i$
colors of the brighter and fainter component for the geometrically- ({\em top})
and kinematically-selected ({\em bottom}) samples of candidate binaries with
$|\delta|<0.4$. The $r-i$ color axes are interpolated from $g-i$ axes using
Eq.~\ref{interp}. Sources are binned in $0.1\times0.1$ mag $g-i$ color pixels
(minimum of 6 sources per pixel), and the median values are color-coded
according to the legends given at the top of each panel. Inset histograms show
the distribution of $\langle \delta \rangle$. The $\langle \delta \rangle$
distribution medians are 0 to within $<0.01$ mag, and the scatter (determined
from the interquartile range) is 0.07 mag for both samples.
\label{medians}}
\end{figure}

\clearpage

\begin{figure}
\epsscale{1.0}
\plotone{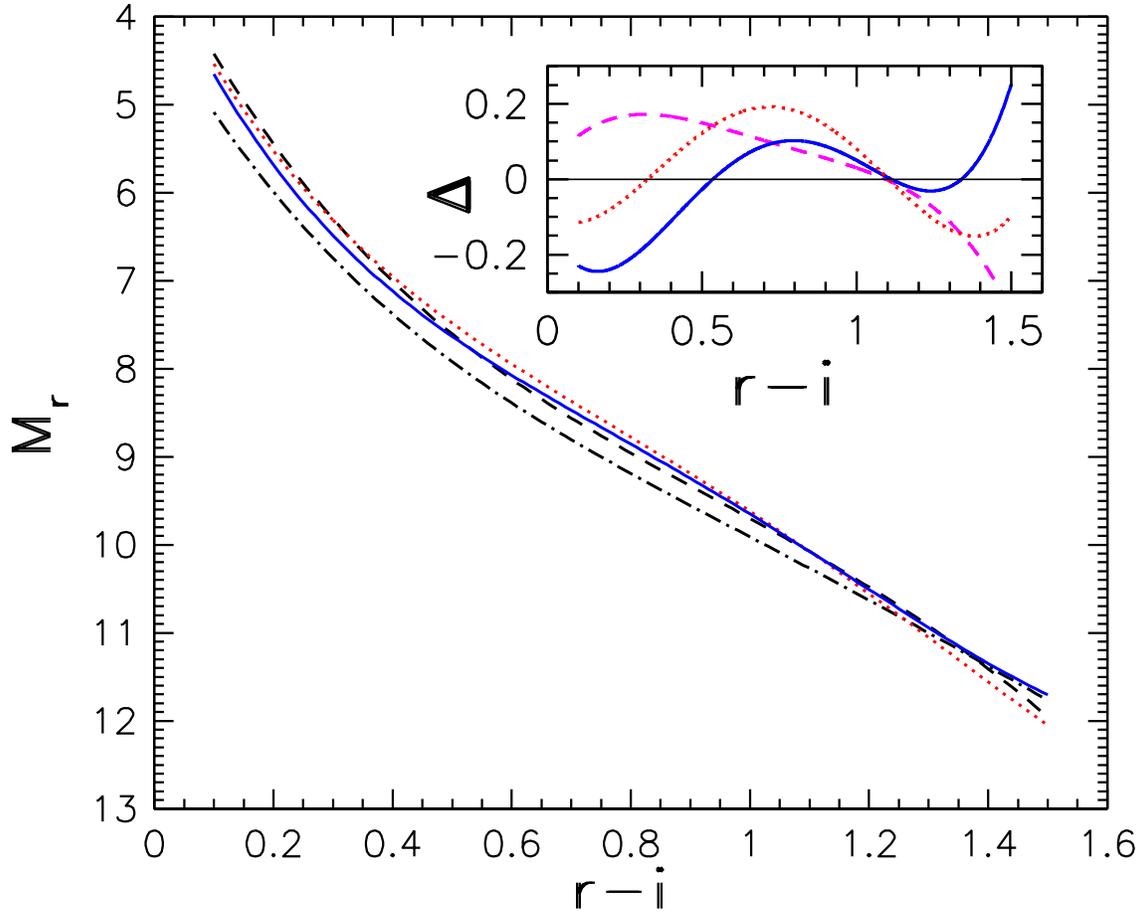}
\caption{
Comparison of Eq.~\ref{Mr_faint} ({\em dot-dashed line}) and Eq.~\ref{Mr_bright}
({\em dashed line}) photometric parallax relations from J08 (their
Eqs.~1 and 2) with Eq.~\ref{Mr_geo} ({\em solid line}) and Eq.~\ref{Mr_kin}
({\em dotted line}) photometric parallax relations determined in this work. The
inset shows the magnitude difference, $\Delta=M_{J08}-M_{S08}$, between the
Eq.~\ref{Mr_bright} photometric parallax relation, and Eqs.~\ref{Mr_geo}
({\em solid line}) and~\ref{Mr_kin} ({\em dotted line}) from this work. The rms
scatter between Eqs.~\ref{Mr_geo} and~\ref{Mr_kin}, and Eq.~\ref{Mr_bright}, is
0.13 mag. The rms scatter between Eqs.~\ref{Mr_geo} and~\ref{Mr_kin}
({\em dashed line}) is also 0.13 mag.
\label{Mr_ri}}
\end{figure}

\clearpage

\begin{figure}
\epsscale{0.5}
\plotone{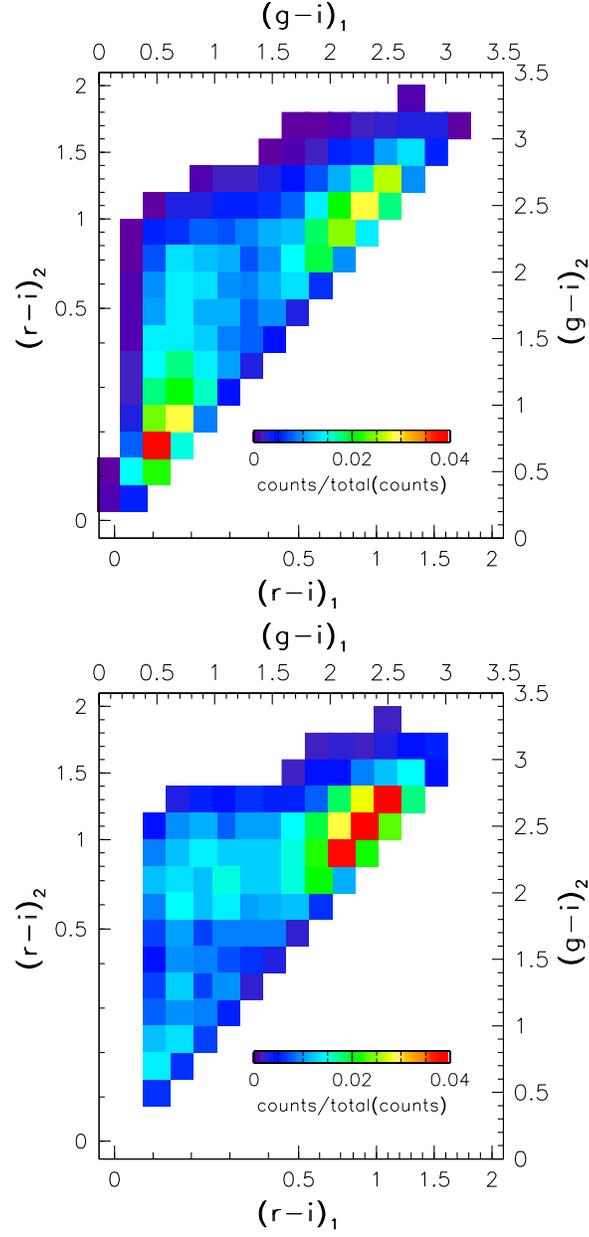}
\caption{
A comparison of $(g-i)_2$ vs.~$(g-i)_1$ color-color distributions of
geometrically-selected ({\em top}) and kinematically-selected disk binaries
({\em bottom}) with $|\delta|<0.4$. The fraction of binaries in a pixel is
color-coded according to legends. The pixels are $0.2\times0.2$ mag wide in
$g-i$ color, and the $r-i$ color axes are interpolated from $g-i$ axes using
Eq.~\ref{interp}.
\label{geo_kin_gi1_gi2_comp}}
\end{figure}

\clearpage

\begin{figure}
\epsscale{0.55}
\plotone{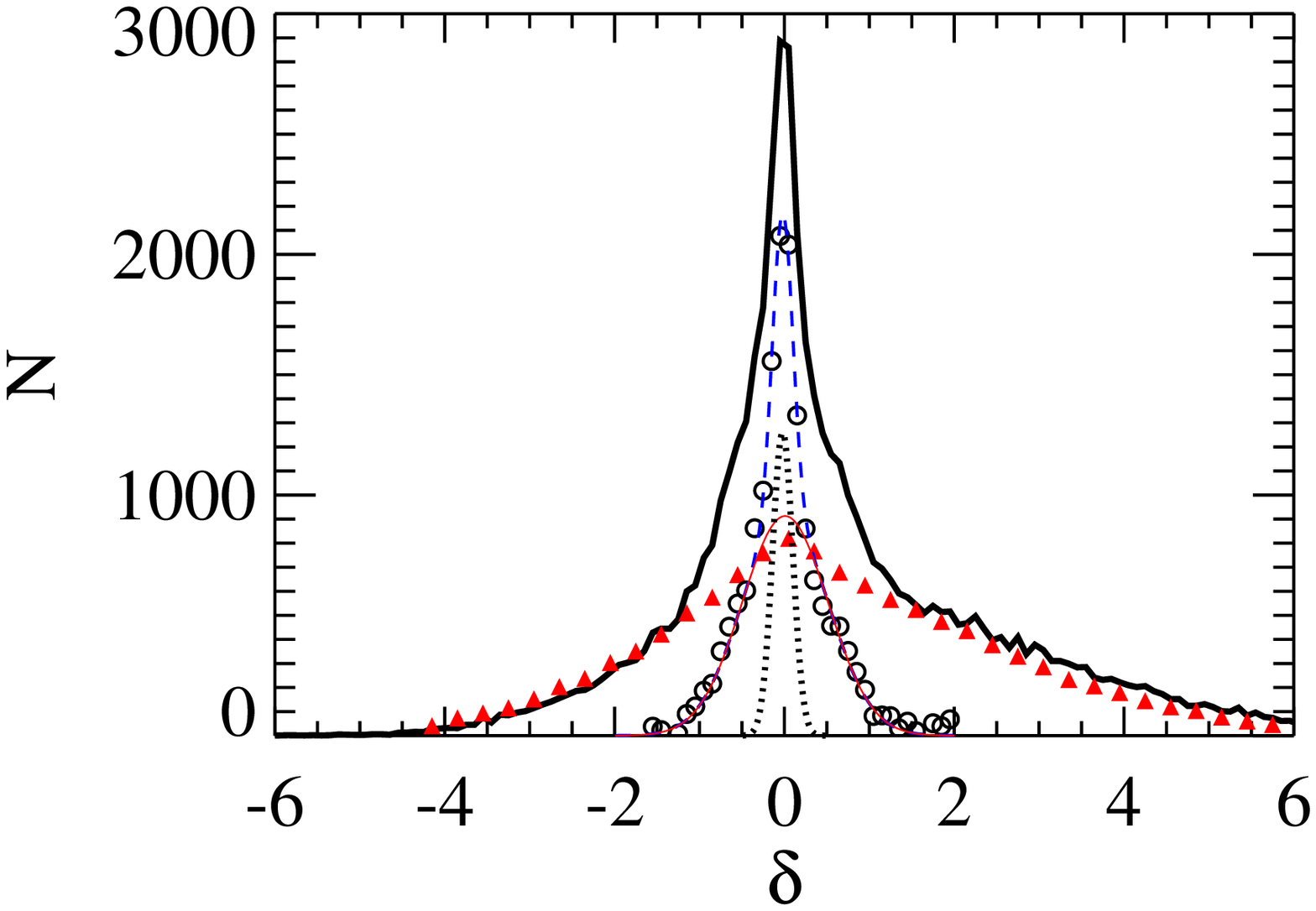}

\plotone{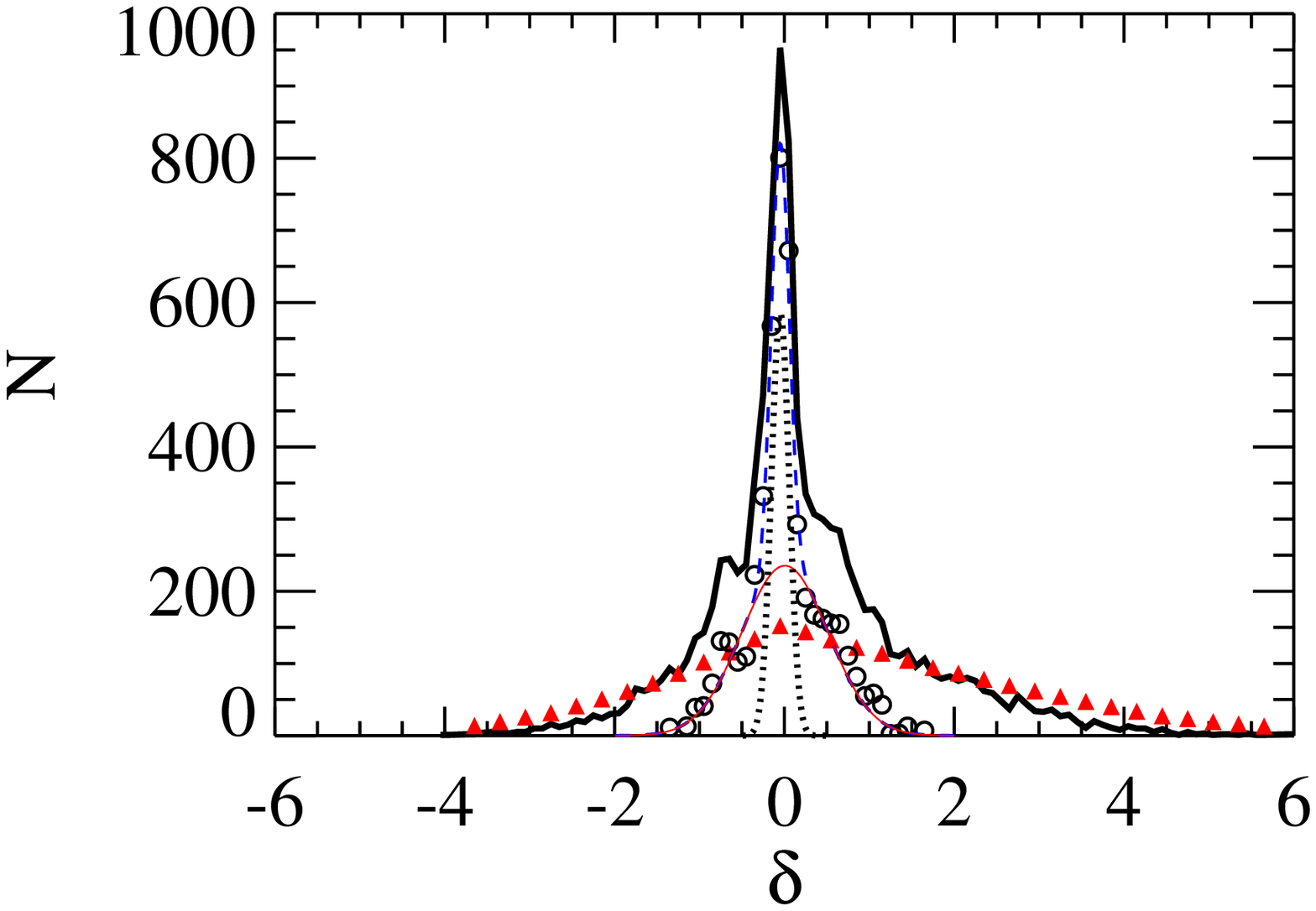}
\caption{
Distribution of $\delta$ values for the geometrically- ({\em top}) and
kinematically-selected ({\em bottom}) samples of candidate binaries, with
absolute magnitudes, $M_r$, calculated using Eqs.~\ref{Mr_geo} and~\ref{Mr_kin},
respectively. The $\delta$ distribution for true binaries ({\em open circles})
is obtained by subtracting the $\delta$ distribution of random pairs
({\em triangles}) from the $\delta$ distribution for candidate
binaries ({\em thick solid line}). The $\delta$ distribution for true binaries
is a non-Gaussian distribution ({\em dashed line}), that can be described as a
sum of two Gaussian distributions. The centers, widths, and areas for the
best-fit narrow ({\em dotted line}) and wide ({\em thin solid line}) Gaussian
distributions are given in Table~\ref{gauss}. The integrals (areas) of $\delta$
distributions for random pairs and candidate binaries are
$A_{random}$ and $A_{observed}$, respectively.
\label{delta_hists}}
\end{figure}

\clearpage

\begin{figure}
\epsscale{0.7}
\plotone{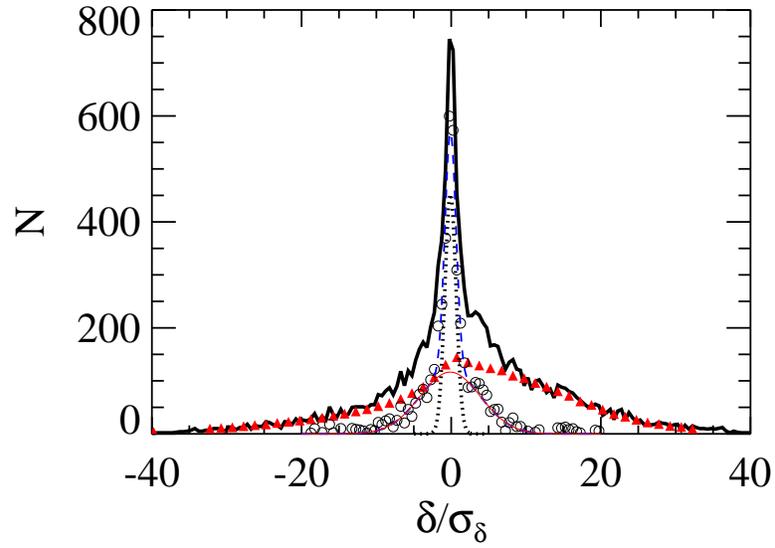}
\caption{
Distribution of $\delta$ values normalized by the expected formal errors,
$\sigma_{\delta}$, for the kinematically-selected sample of candidate binaries.
The $\delta/\sigma_{\delta}$ distribution for true binaries ({\em open circles})
is obtained by subtracting the $\delta/\sigma_{\delta}$ distribution of random
pairs ({\em triangles}) from the $\delta/\sigma_{\delta}$ distribution for
candidate binaries ({\em thick solid line}). The $\delta/\sigma_{\delta}$
distribution for true binaries is a non-Gaussian distribution ({\em dashed
line}), that can be described as a sum of two Gaussian distributions. The
best-fit narrow Gaussian ({\em dotted line}) is 0.75 wide and centered on -0.10,
while the best-fit wide Gaussian ({\em thin solid line}) is 4.04 wide and
centered on -0.14.
\label{delta_norm}}
\end{figure}

\clearpage

\begin{figure}
\epsscale{1.0}
\plotone{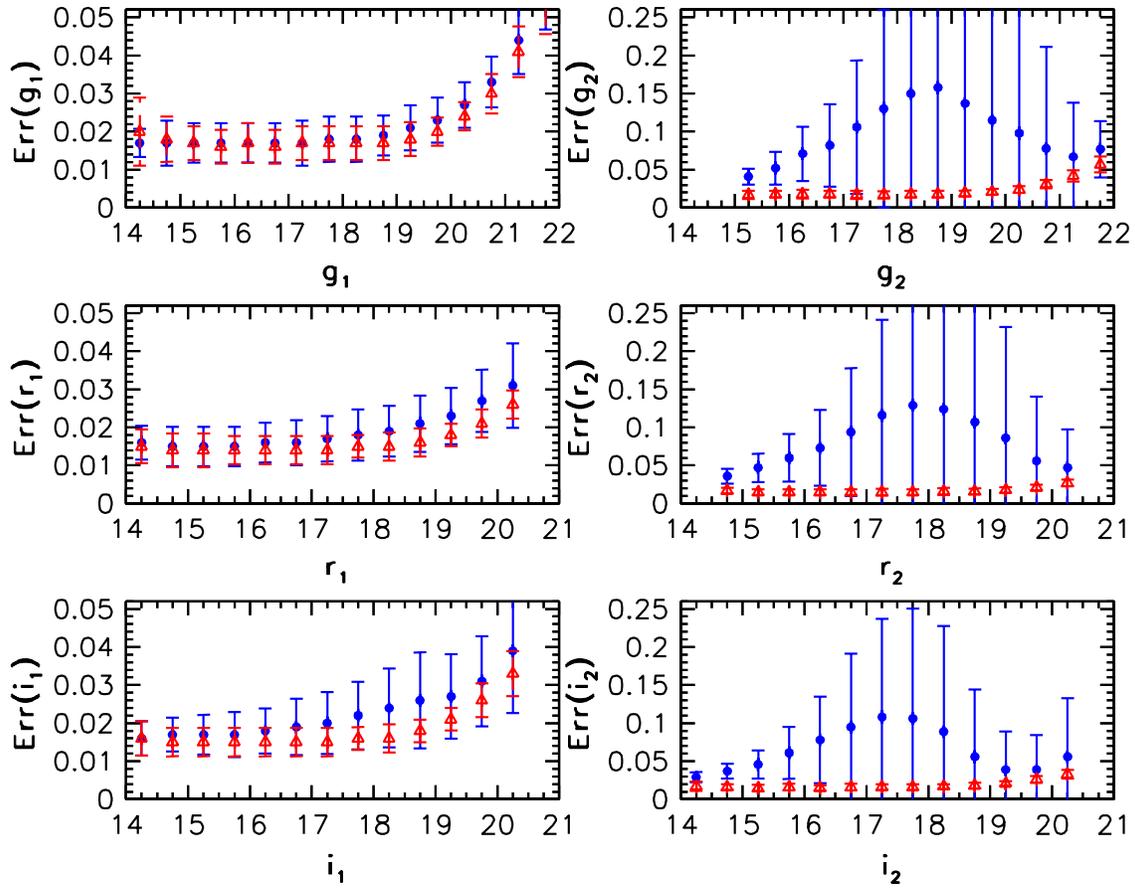}
\caption{
Dependence of median PSF magnitude errors on magnitude for the brighter
({\em left}) and fainter ({\em right}) components in the geometrically-
({\em dots}) and kinematically-selected ({\em triangles}) samples of candidate
binaries. The vertical bars show the rms scatter in each bin (not the error of
the median which is much smaller). The fainter components of
geometrically-selected candidate binaries have overestimated median PSF
magnitude errors when compared to the kinematically-selected binaries.
\label{photo_errors}}
\end{figure}

\clearpage

\begin{figure}
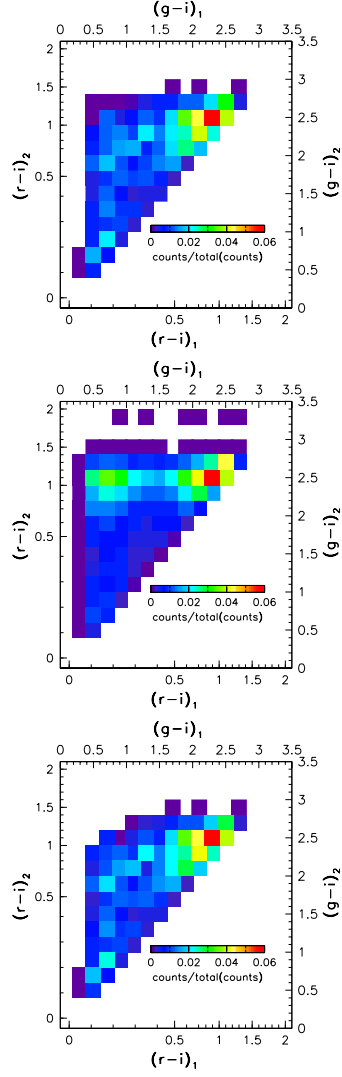

\epsscale{0.29}
\plotone{f15a.eps}

\plotone{f15b.eps}

\plotone{f15c.eps}
\caption{
The fraction of $|\delta|<0.4$ binaries in $0.7 < d/kpc < 1.0$ volume-complete
geometrically-selected ({\em top}) and random ({\em middle}) samples that have
$(g-i)_1$ and $(g-i)_2$ as the colors of the brighter and fainter component.
The pixels are $0.2\times0.2$ mag wide in $g-i$ color, and the $r-i$ color axes
are interpolated from $g-i$ axes using Eq.~\ref{interp}. The pixels in maps sum
to 1. The bottom plot shows the difference,
$f_{cand}[(g-i)_1,(g-i)_2]-C*f_{rand}[(g-i)_1,(g-i)_2]$, between the two maps,
where $C=0.14$ is the fraction of random pairs estimated
using Eq.~\ref{eff} for the $|\delta|<0.4$, $0.7 < d/kpc < 1.0$
geometrically-selected sample. The pixels with negative values are not shown and
the map is renormalized so that the pixels sum to 1.
\label{counts_0.7_1.0}}
\end{figure}

\clearpage

\begin{figure}
\epsscale{1.0}
\plotone{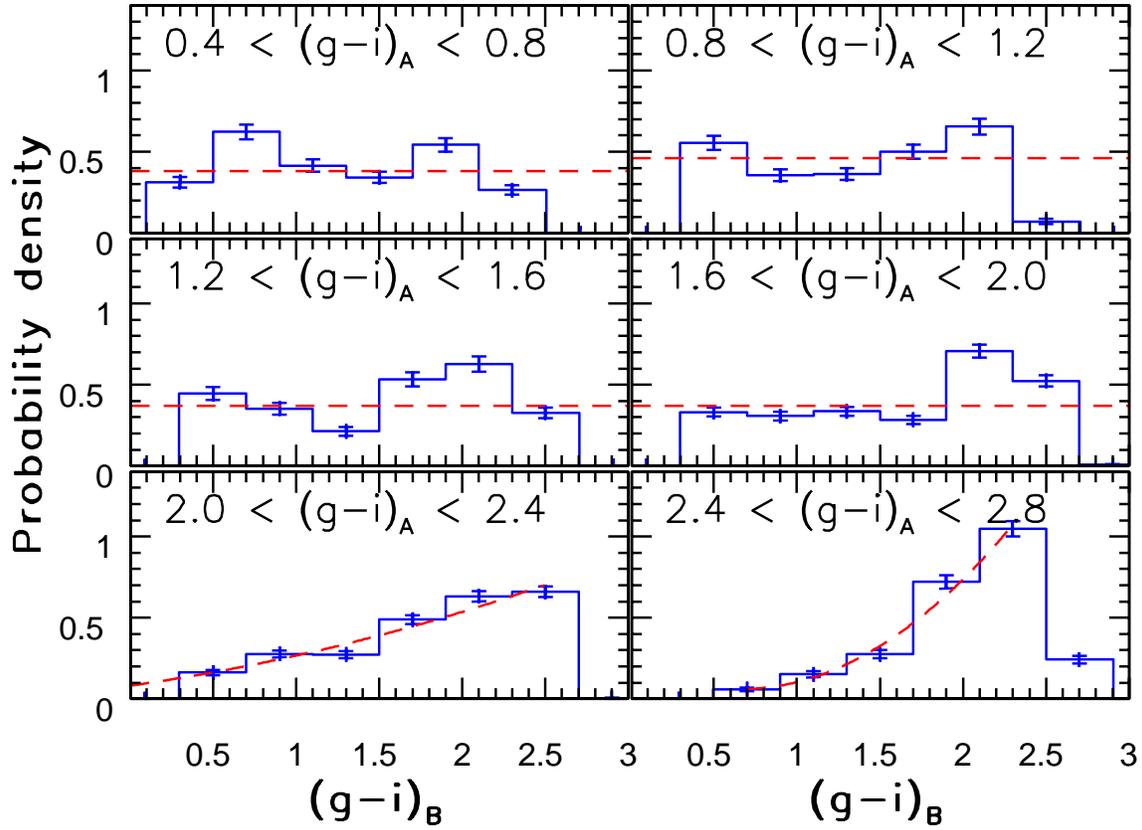}
\caption{
Conditional probability density of having one component with $(g-i)_B$ color in
a wide binary system where the other component has $(g-i)_A$. The conditional
probability density for $(g-i)_A < 2.0$ ({\em top} and {\em middle}) is
independent of $(g-i)_B$, while for $(g-i)_A > 2.0$ ({\em bottom}) it changes as
a square of $(g-i)_B$. The best-fit functions describing these conditional
probabilities are given in Table~\ref{tbl_wide_bin_prob}.
\label{wide_binary_prob}}
\end{figure}

\clearpage

\begin{figure}
\epsscale{0.5}
\plotone{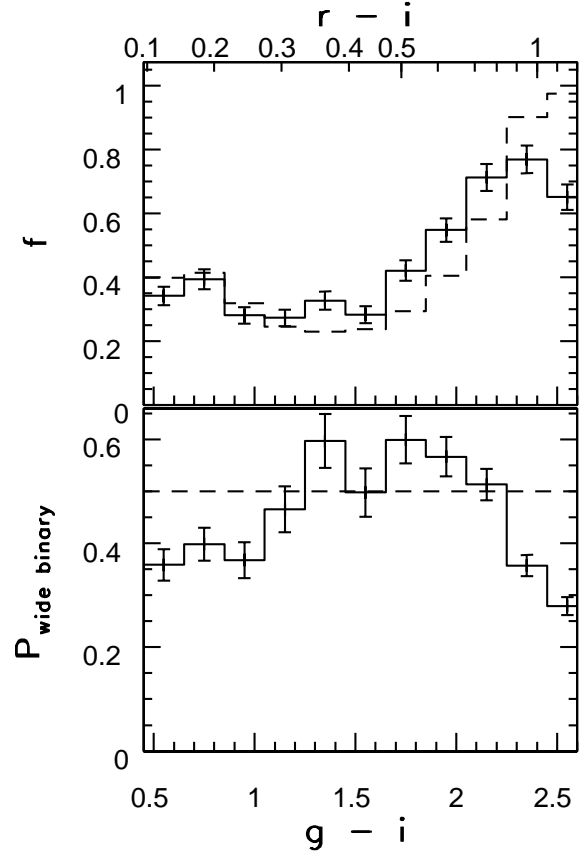}
\caption{
{\em Top:} A comparison of $g-i$ color distribution of stars in the
$|\delta|<0.4$, $0.7 < d/kpc < 1.0$ volume-complete, geometrically-selected,
wide binary sample ({\em solid line}), and of all stars in the same volume
({\em dashed line}). The distributions are normalized to an area of 1, and the
error bars show the Poisson noise. {\em Bottom:} The probability density for
finding a star with $g-i$ color in a wide binary system,
$P[(g-i)_A] = P_{wide binary}$, calculated as a ratio of the two distributions
from the top panel, and renormalized to an area of 1. The equal probability
distribution is shown as the dashed line. 
\label{prob_gi}}
\end{figure}

\clearpage

\begin{figure}
\epsscale{0.38}
\plotone{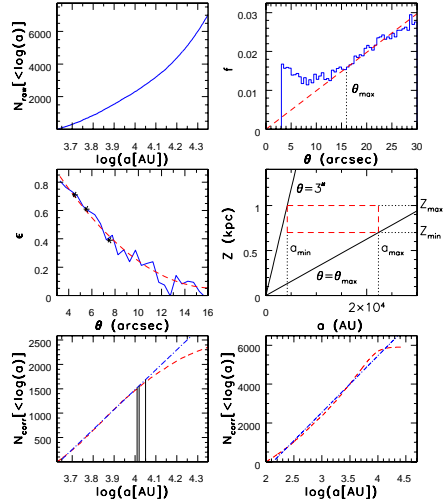}
\caption{
{\em Top left:} The cumulative distribution of $\log(a)$ for
geometrically-selected candidate binaries with $|\delta|<0.2$ and
$0.7< Z/kpc < 1.0$, where $a$ is the average semi-major axis. {\em Top right:}
The differential distribution of angular separation, $\theta$, for
geometrically-selected candidate binaries with $|\delta|<0.2$ and
$0.7< Z/kpc < 1.0$. The distribution of random pairs ({\em dashed line}) is
obtained by fitting a linear function $f_{rnd}(\theta) = C \, \theta$ to the
observed histogram for $\theta > 18\arcsec$. $\theta_{max}$ is defined as the
angular separation for which the fraction of true binaries falls below
$\sim5\%$. {\em Middle left:} The fraction of true binaries, $\epsilon$
({\em solid line}), calculated from the $\theta$ distribution using
Eq.~\ref{eff2} (see Section~\ref{geo_select}) for the $0.7< Z/kpc < 1.0$ sample,
is modeled as a second-degree polynomial, $\epsilon(\theta)$
({\em dashed line}). For three $\theta-$selected subsamples
($4\arcsec-5\arcsec$, $5\arcsec-6\arcsec$, and $7\arcsec-8\arcsec$), the
fraction of true binaries was also calculated using Eq.~\ref{eff} (i.e., from
the $\delta$ distribution) and is shown with symbols. {\em Middle right:} The
box ({\em dashed lines}) shows the allowed range in $a$ defined by $Z_{min}$,
$Z_{max}$, and $\theta_{max}$ (see Eqs.~\ref{a_min} and~\ref{a_max}). Only
binaries within this $a$ range are considered when plotting the corrected
cumulative distribution of $\log(a)$. {\em Bottom left:} The cumulative
distribution of $\log(a)$ for candidate binaries with $|\delta|<0.2$ and
$0.7< Z/kpc < 1.0$ ({\em dashed line}), corrected using $\epsilon(\theta)$ to
account for the decreasing fraction of true binaries at large
$\theta\propto a/d$ separations. The vertical lines show $\log(a)$ for which the
straight line fit ({\em dot-dashed line}) to the cumulative distribution
deviates by more than $1.0\%$ ($\log(a_{low})$), $1.5\%$ ($\log(a_{break})$),
and $2.0\%$ ($\log(a_{high})$). {\em Bottom right:} The corrected cumulative
distribution of $\log(a)$ for mock candidate binaries created using the
$f(a)\propto a^{-0.8}$ distribution limited to $a_1 = 100$ AU and $a_2 = 10000$
AU.
\label{loga_cum}}
\end{figure}

\clearpage

\begin{figure}
\epsscale{0.9}
\plotone{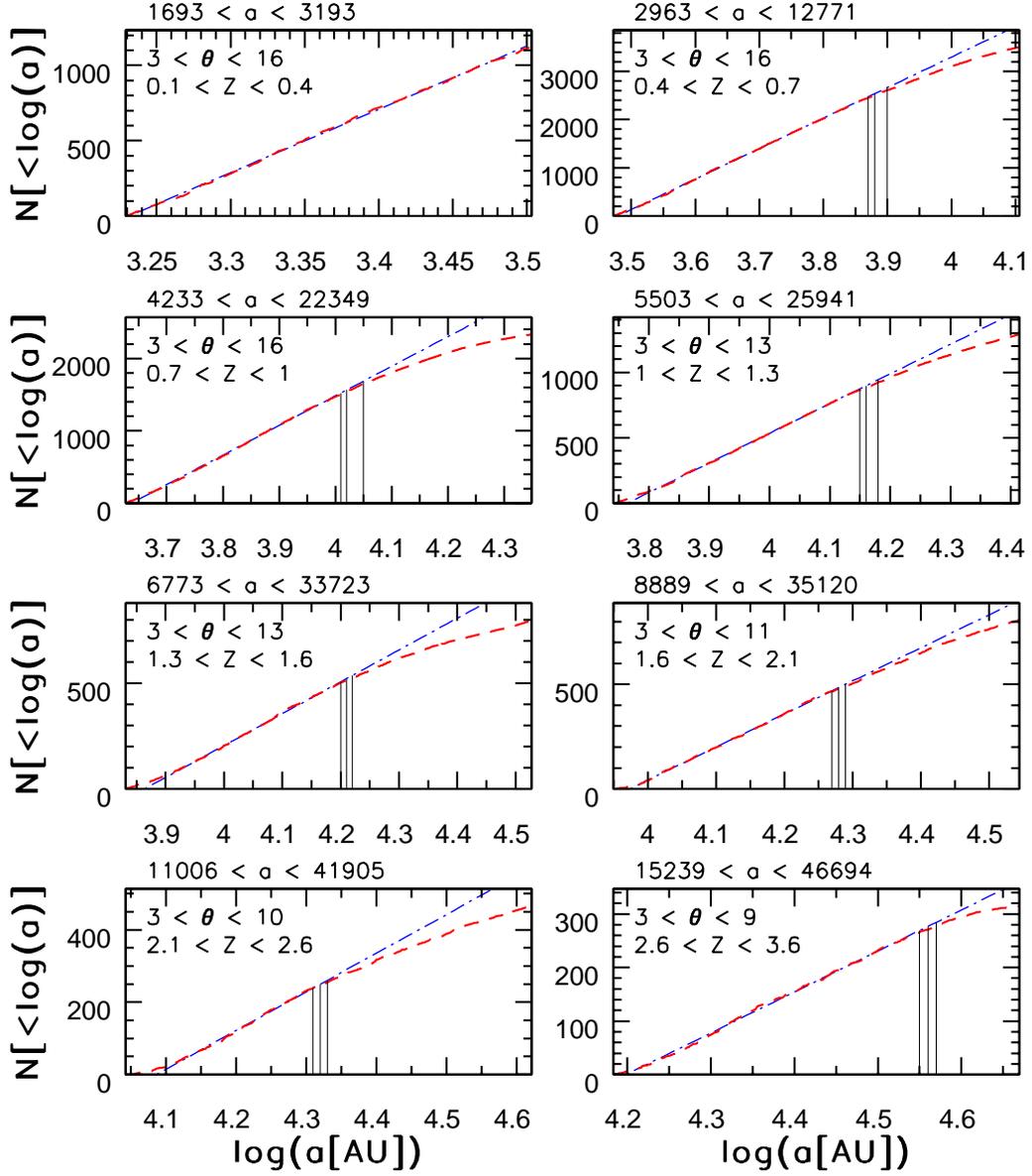}
\caption{
Similar to Fig.~\ref{loga_cum} ({\em bottom}) plot, but for different $Z$
(height above the Galactic plane) bins ranging from $0.1< Z/kpc < 0.4$
({\em top left}) to $2.6< Z/kpc < 3.6$ ({\em bottom right}). The sampled range
of average semi-major axes and angular separations is given for each panel. In
the $0.1< Z/kpc < 0.4$ bin ({\em top left}), the upper limit on
$\log(a_{break})$ is 3.50.
\label{loga_cum_fits}}
\end{figure}

\clearpage

\begin{figure}
\epsscale{1.0}
\plotone{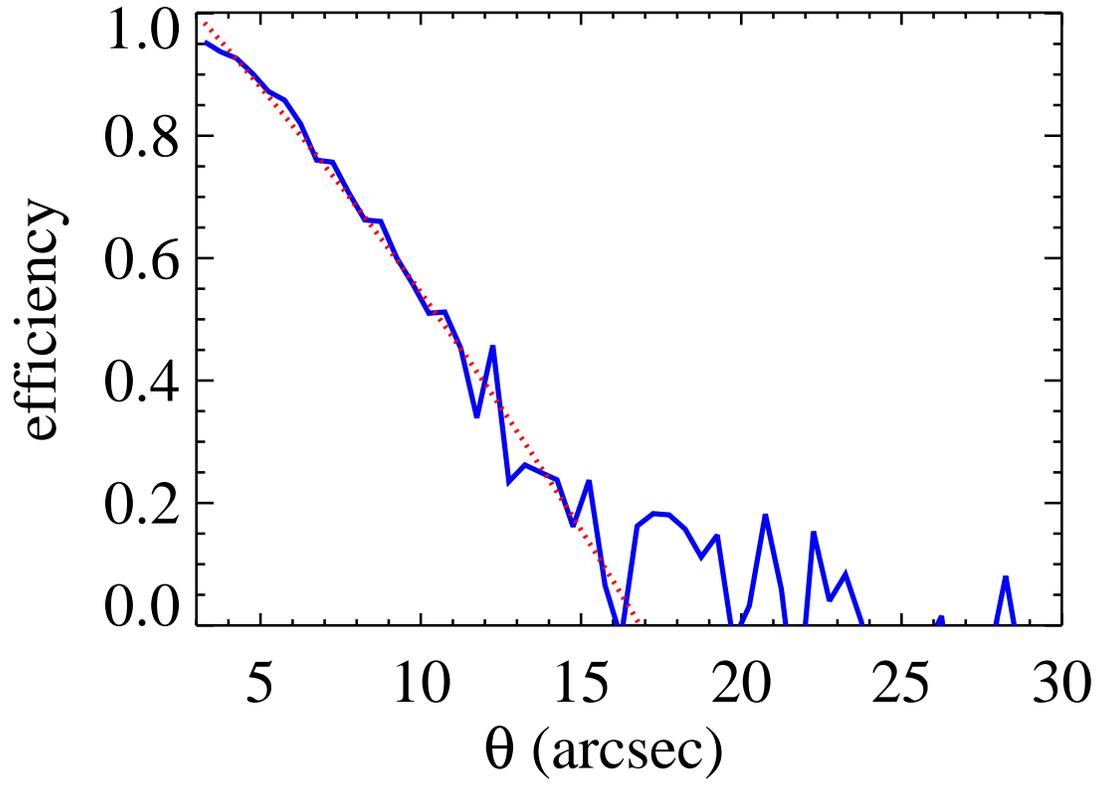}
\caption{
The fraction of true binaries ($\epsilon$) in the $0.1 <Z/kpc < 0.4$,
$|\delta|<0.2$ geometrically-selected sample as a function of angular
separation. The fraction goes below $\sim5\%$ at $\theta_{max}=16\arcsec$, and
puts the upper limit on probed semi-major axes to $\sim3,200$ AU.
\label{theta_eff_0.1_0.4_geo}}
\end{figure}

\clearpage

\begin{figure}
\epsscale{0.7}
\plotone{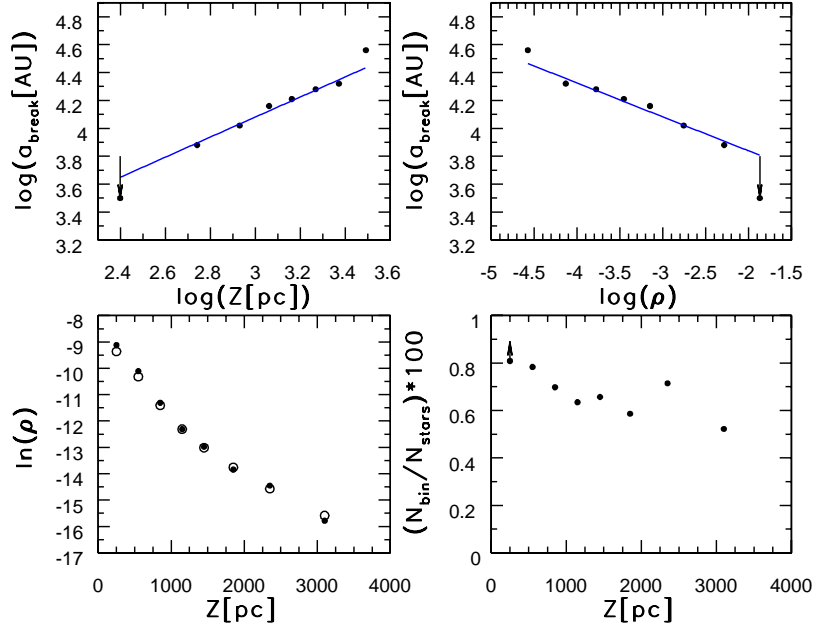}
\caption{
{\em Top left:} The dependence of $\log(a_{break})$ values
(c.f.~Fig.~\ref{loga_cum_fits}) on $\log(Z)$ ({\em dots}) is modeled as
$\log(a_{break})=k \, \log(Z)+l$, where $k=0.72\pm0.05$ and $l=1.93\pm0.15$,
or approximately, $a_{break}[AU] = 12,302 \, Z[kpc]^{0.72}$. The symbol size
shows the range between $\log(a_{low})$ and $\log(a_{high})$. The arrow
indicates that the $\log(a_{break})$ in the $0.1< Z/kpc < 0.4$ bin
($\log(Z)\sim2.4$) is an upper limit. {\em Top right:} The dependence of
$\log(a_{break})$ on $\log(\rho)$, where $\rho$ is the local number density of
stars, is modeled as $\log(a_{break})=k \, \log(\rho)+l$, where $k=-0.24\pm0.02$
and $l=3.35\pm0.07$, or $a_{break} \propto \rho^{-1/4}$. {\em Bottom left:} The
dependence of local number density, $\ln(\rho)$, of binaries ({\em dots}) and
stars ({\em circles}) on the height above the Galactic plane, where the density
of stars is normalized to match the density of binaries at 1 kpc.
{\em Bottom right:} The fraction of binaries relative to the total number of
stars as a function of the height above the Galactic plane. The arrow shows the
predicted fraction of binaries in the $0.1< Z/kpc < 0.4$ bin, if the $a_{break}$
value follows the $a_{break}\propto Z^{0.72}$ relation.
\label{logZ_plots}}
\end{figure}

\clearpage

\begin{figure}
\epsscale{1.0}
\plotone{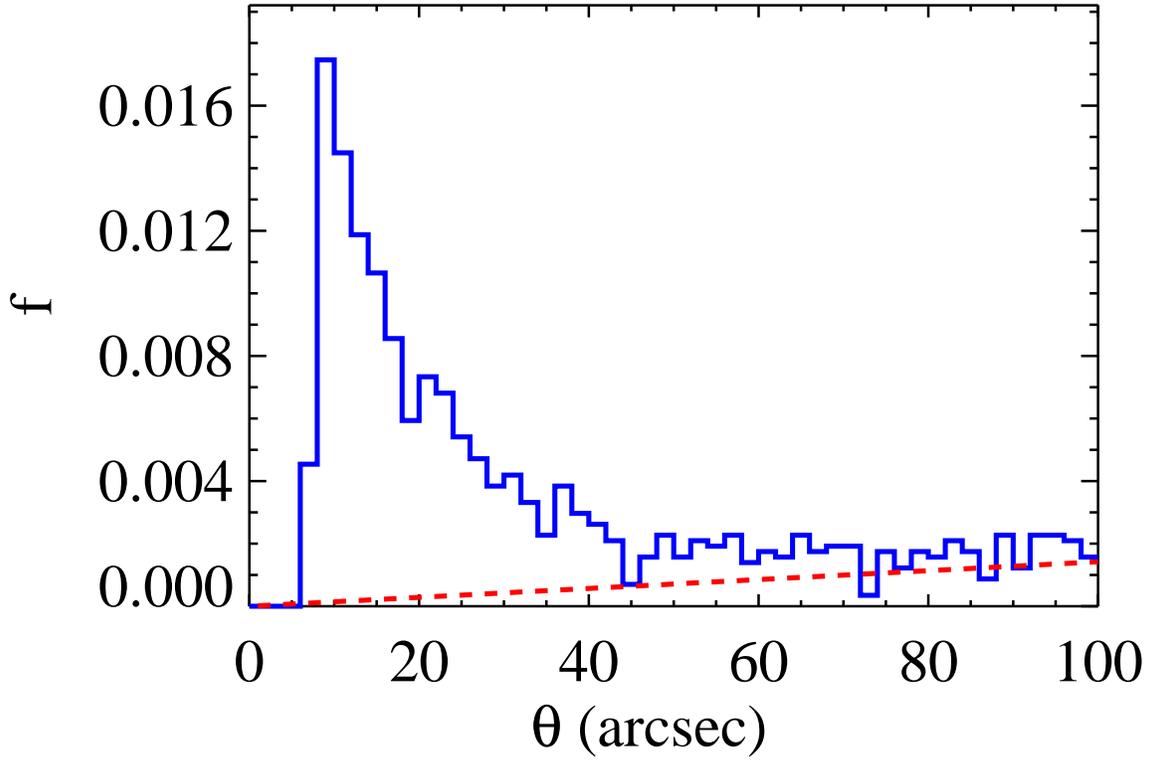}
\caption{
The distribution of angular separation for the $0.1 < Z/kpc < 0.4$,
$|\delta|<0.2$ kinematically-selected sample of candidate binaries. The data
({\em solid line}) extend to $\theta=500\arcsec$, though the plotted range is
restricted for clarity. The distribution of random pairs ({\em dashed line})
was obtained by fitting $f_{rnd}(\theta) = C \, \theta$ to the observed
histogram for $\theta > 200\arcsec$. The sharp drop-off in the observed
distribution for $\theta\la9$ is probably due to blending of close pairs in the
POSS data.
\label{theta_fit_500pc}}
\end{figure}

\clearpage

\begin{figure}
\epsscale{1.0}
\plotone{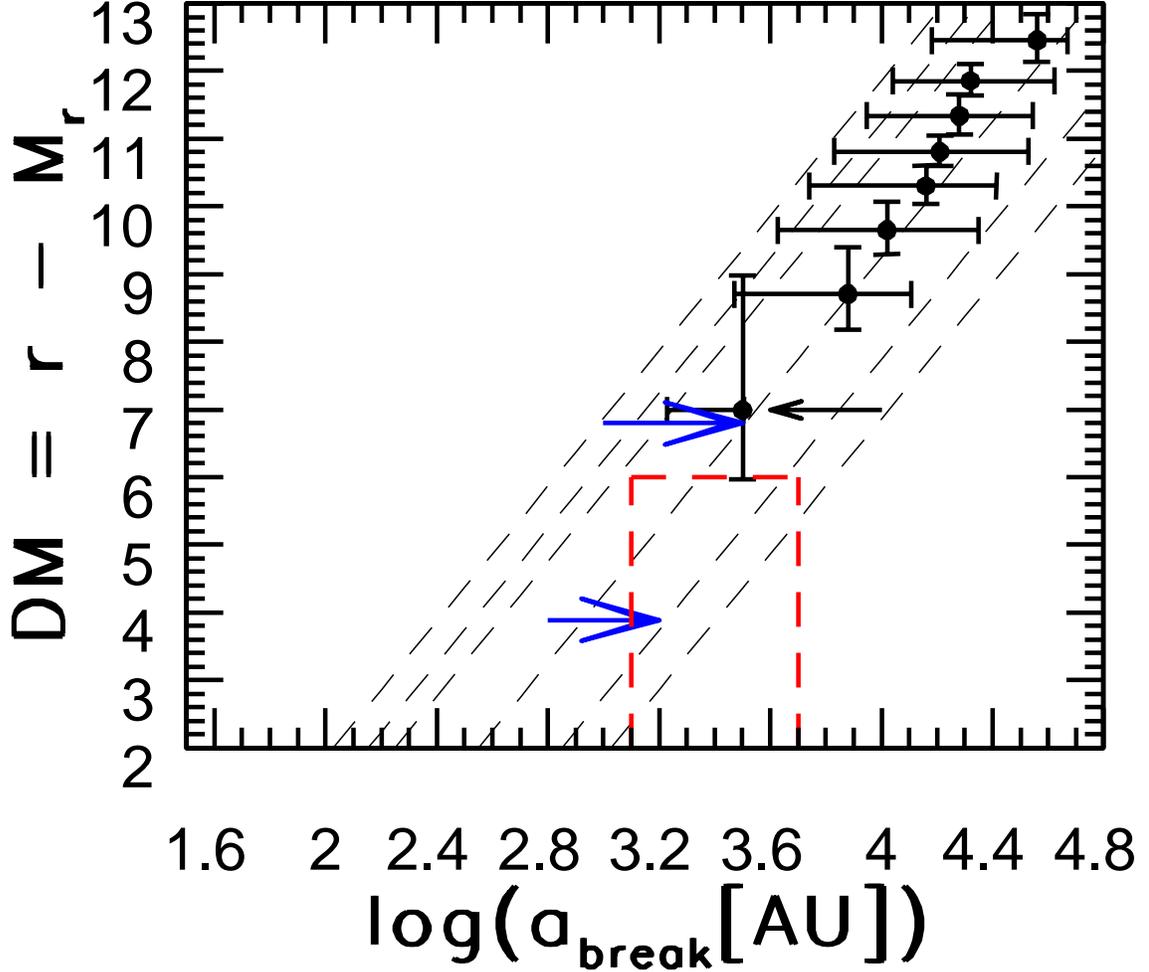}
\caption{A comparison of results for the turnover in the distribution of
semi-major axes, $a_{break}$, as a function of distance modulus, of wide binary
systems determined here (symbols with error bars; the horizontal bars mark the
range of probed semi-major axes, and the vertical bars mark the width of the
distance bins; the lowest point is only a lower limit, for the sake of
comparison with other results we ignore the difference between distance from us
and distance from the Galactic plane because our sample is dominated by high
galactic latitude stars), determined by L\'{e}pine \& Bongiorno (2007; the
dashed rectangle indicates  constraint on $a_{break}$ and the probed distance
range), and determined by Chanam\'{e} \& Gould (2004; big arrows, indicating
upper limits on $a_{break}$ and the probed distance range; the point at larger
distance modulus corresponds to halo binaries). The diagonal dashed lines are
lines of constant angular scale, $\theta$, for values of $3\arcsec$, $4\arcsec$,
$5\arcsec$, $10\arcsec$, $20\arcsec$ and $30\arcsec$ (from left to right).
\label{Fig:compare}}
\end{figure}

\clearpage

\begin{figure}
\epsscale{0.7}
\plotone{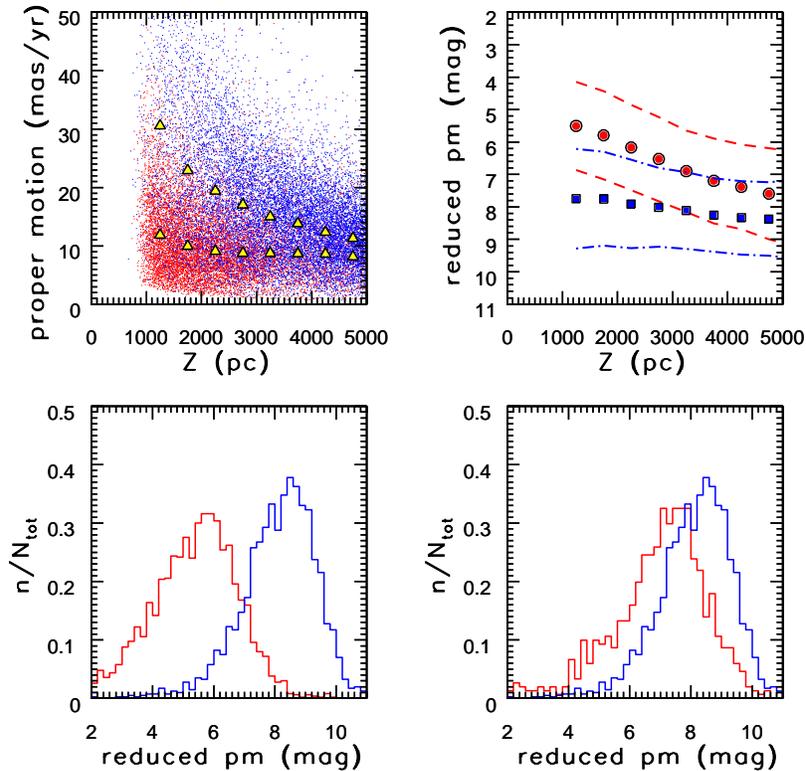}
\caption{The top left panel shows the proper motion distribution as a function
of distance from the Galactic plane ($Z$) for a sample of $\sim$16,000 likely
disk stars (red dots) and a sample of $\sim$34,400 likely halo stars (blue
dots). All stars have $14<r<20$ and $0.2<g-r<0.4$, and are separated using
photometric metallicity. The triangles show the median values in 500 pc wide $Z$
bins for each sample (lower symbols: disk, upper symbols: halo). Note that the
median proper motion for disk stars becomes constant beyond $Z\sim2$ kpc due to
the vertical gradient of rotational velocity for disk stars. The top right panel
shows the median position (symbols) and widths (lines; $\pm1\sigma$ envelope
around the medians) of the reduced proper motion sequences for disk (red dots
and dashed lines) and halo (blue squares and dot-dashed lines) stars, as
functions of $Z$. The two bottom panels show the cross-sections of the reduced
proper motion sequences for stars with $Z=1-1.5$ kpc (bottom left; red histogram
for disk stars and blue for halo stars) and $Z=3.5-4$ kpc. The histograms are
normalized by the total number of stars in each subsample. The disk-to-halo star
count ratio is 4.3 in the left panel, and 0.38 in the right panel. Note the
significant overlap of the two sequences for large $Z$.
\label{Fig:App1}}
\end{figure}

\clearpage

\begin{figure}
\epsscale{0.8}
\plotone{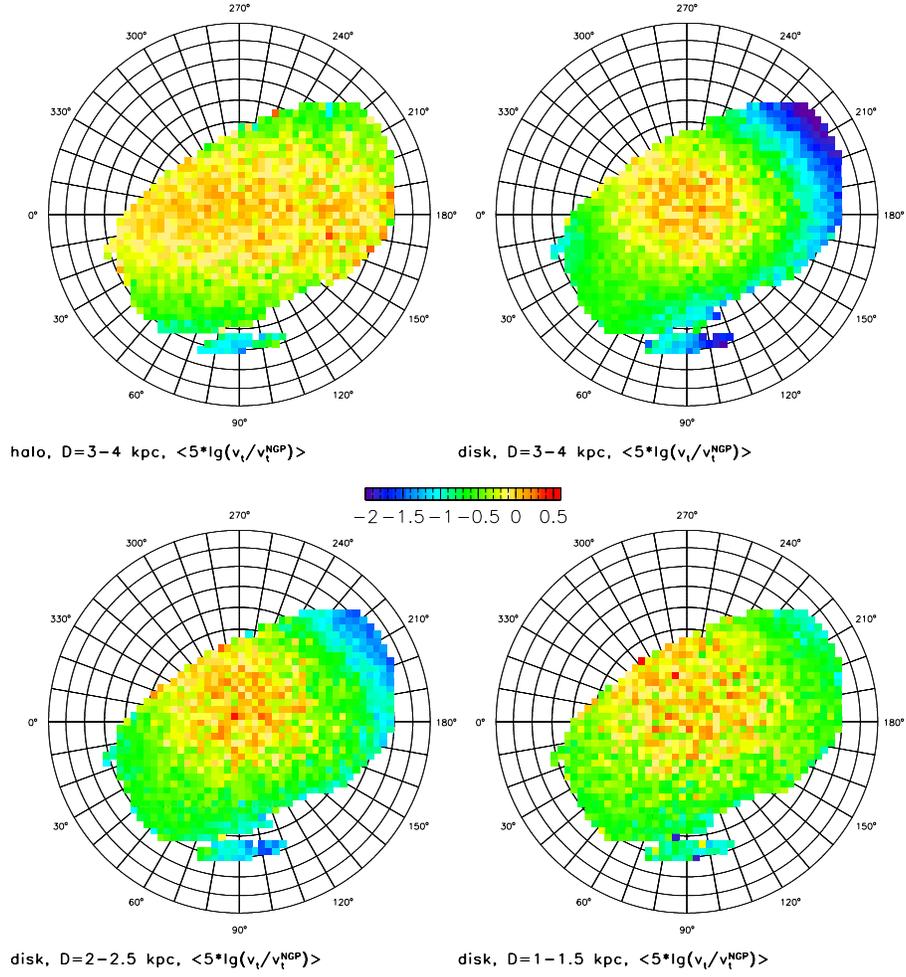}
\caption{An illustration of the offsets in the position of reduced proper motion
sequences as a function of distance, position on the sky and population. Each
panel shows the median value of $5\log(v_t/v_t^{NGP})$, where $v_t$ is the
heliocentric tangential velocity, and $v_t^{NGP}$ is its value at the north
Galactic pole, in Lambert projection of northern galactic hemisphere. The maps
are color-coded according to the legend shown in the middle of the figure
(magnitudes), and are constructed using stars with $0.2<g-r<0.4$. Stars are 
separated into halo and disk populations using photometric metallicity (for
details see I08a). The top left panel shows results for halo stars with
distances in the 3-4 kpc range. The other three panels correspond to disk stars
in the distance range 3-4 kpc, 2-2.5 kpc, and 1-1.5 kpc.}
\label{Fig:App2}
\end{figure}

\clearpage

\begin{figure}
\epsscale{0.7}
\plotone{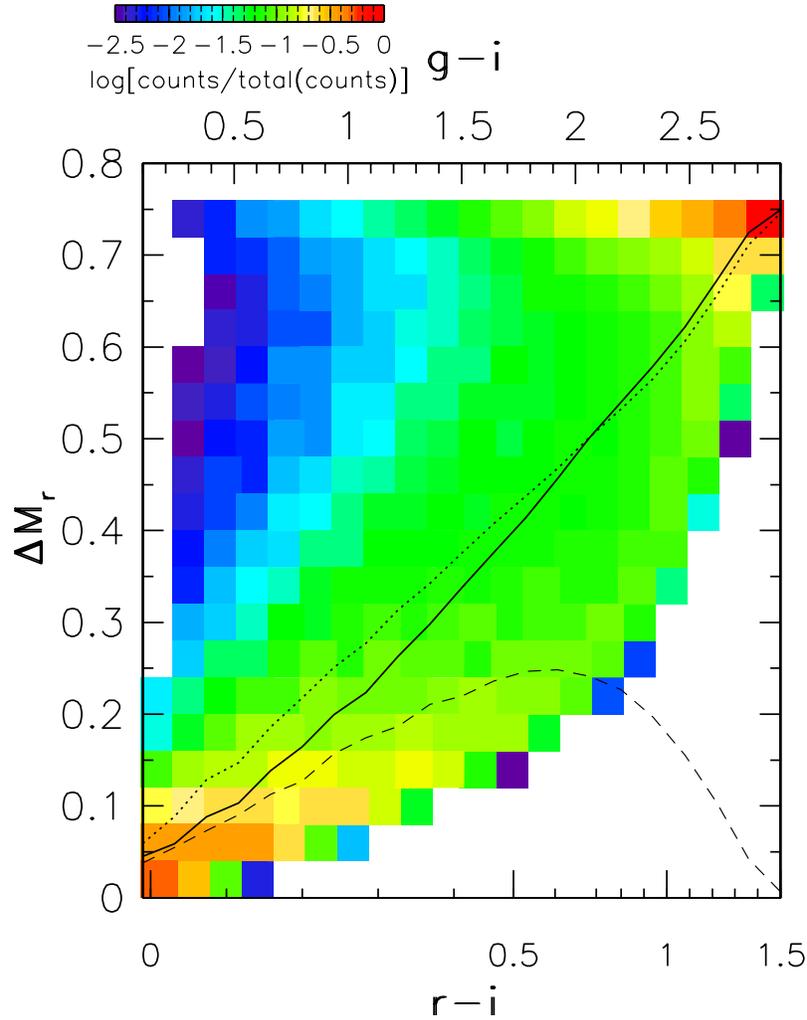}
\caption{
The number of unresolved binary systems (normalized with the total count in a
given $g-i$ bin) with a magnitude offset $\Delta M_r=M_r(assumed)-M_r(true)$ as
a function of the system's $g-i$ color. The assumed absolute magnitude for a
system with a $g-i$ color, $M_r(assumed)$, was calculated using
Eq.~\ref{Mr_bright} (Eq.~1 from J08), while the true absolute magnitude,
$M_r(true)$ was calculated by adding up luminosities of components. The mean,
median, and the rms scatter of $\Delta M_r$ are shown with the dotted, solid,
and dashed lines, respectively.
\label{dMr_gi}}
\end{figure}

\clearpage

\begin{figure}
\epsscale{0.7}
\plotone{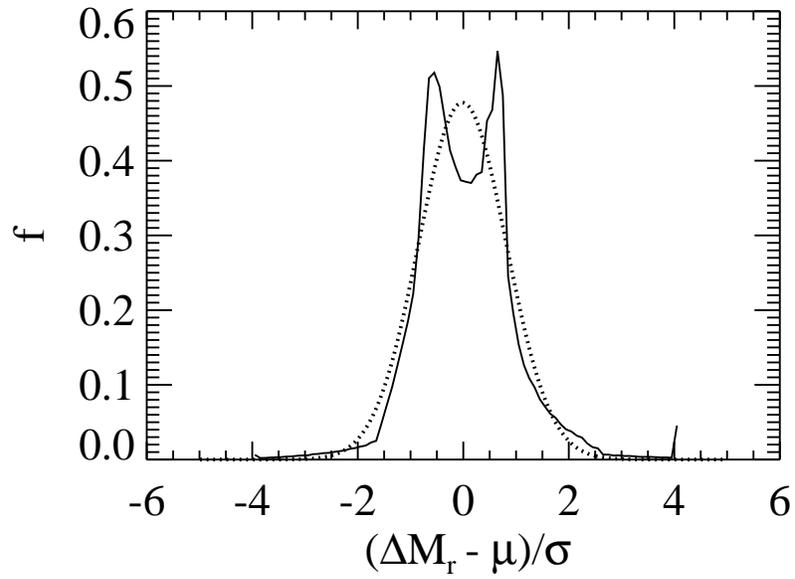}
\caption{
Distribution of differences between the magnitude offset, $\Delta M_r$, and the
median magnitude offset, $\mu$, normalized with rms scatter, $\sigma$,
({\em solid line}) can be modeled as a 0.9 wide Gaussian ({\em dotted line}).
\label{dMr_chi_gauss}}.
\end{figure}

\clearpage

\begin{figure}
\epsscale{1.0}
\plotone{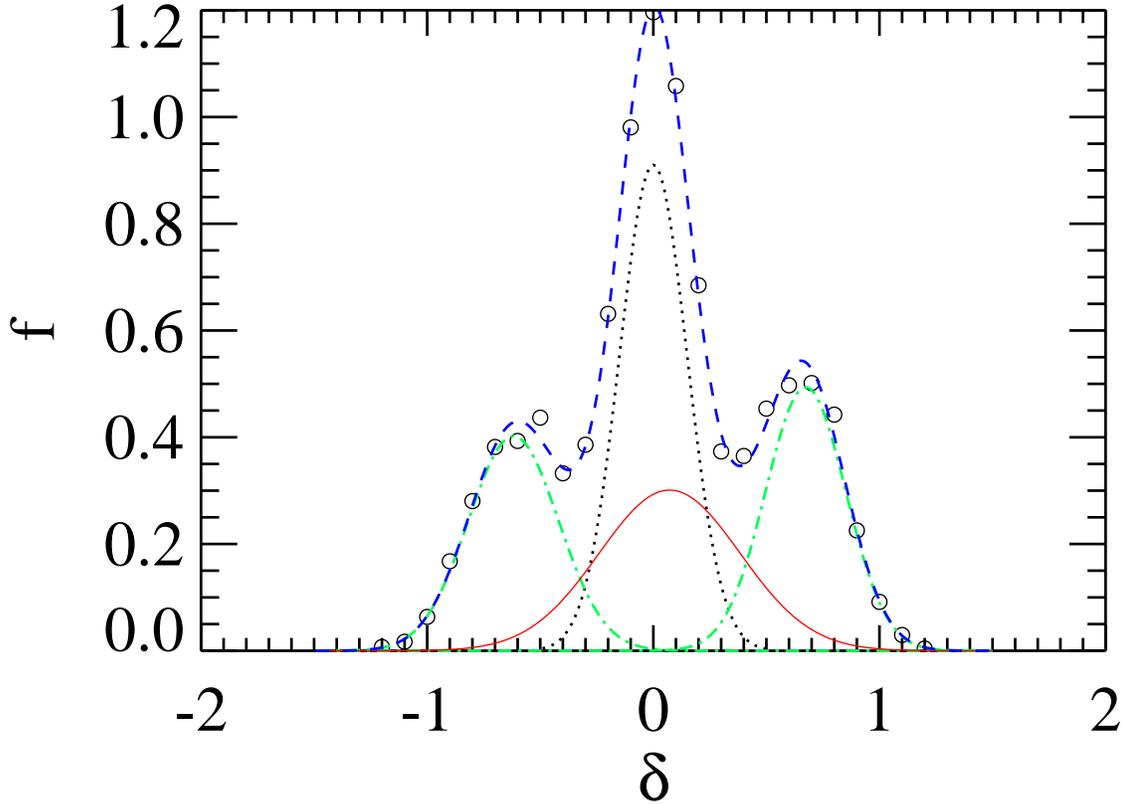}
\caption{
The distribution of $\delta$ values for the mock sample of wide binaries with
both components redder than $g-i=2.0$ ({\em open circles}). In this sample, a
star has a 40\% probability to be an unresolved binary system. Single
star-single star configurations contribute the central narrow Gaussian
({\em dotted line}), unresolved binary-unresolved binary configurations
contribute the central wide Gaussian ({\em thin solid line}), while the single
star-unresolved binary configurations contribute the left and the right
Gaussians ({\em dot-dashed lines}). The centers, widths, and areas of Gaussians
are: $N_1(0.00,0.15,0.34)$, $N_2(0.06,0.35,0.28)$, $N_3(-0.64,0.18,0.18)$,
$N_4(0.71,0.17,0.19)$ for the narrow, wide, left, and right Gaussians,
respectively.
\label{gauss_fit_red}}
\end{figure}

\end{document}